\documentclass[12pt]{article}
\usepackage{amssymb}

\usepackage{amsmath,amscd,amsthm}

\newtheorem{theorem}{Theorem}[section]

\newtheorem{lemma}[theorem]{Lemma}

\newtheorem{remark}[theorem]{Remark}

\begin{document}

\title{On general Cwikel-Lieb-Rozenblum and Lieb-Thirring inequalities}
\author{S. Molchanov, B. Vainberg \thanks{
The authors were partially supported by the NSF grant DMS-0706928.} \and %
Dept. of Mathematics, University of North Carolina at Charlotte, \and %
Charlotte, NC 28223, USA \\[5mm]
\hfill \hspace{70mm} {\it \normalsize To our dear friend Vladimir
Maz'ya} \\[-10mm]}

%\dedicatory{To our dear friend Vladimir Maz'ya}
\date{}

\maketitle

\begin{abstract}
These classical inequalities allow one to estimate the number of negative eigenvalues
and the sums $S_{\gamma}=\sum |\lambda_i|^{\gamma}$ for a wide class of
Schr\"{o}dinger operators. We provide a detailed proof of these inequalities
for operators on functions in metric spaces using the classical Lieb approach based on the Kac-Feynman formula. The main goal of the paper is a new set of examples
which include perturbations of the
Anderson operator, operators on free, nilpotent and
solvable groups, operators on quantum graphs,
Markov processes with independent increments. The study of the examples requires an
 exact estimate of the kernel of the corresponding parabolic
semigroup on the diagonal. In some cases the kernel decays exponentially as $t\rightarrow \infty $.
This allows us to consider very slow decaying potentials and obtain some
results that are precise in the logarithmical scale.

\textbf{MSC: }35P15, 47A75, 47B99, 20P05, 60J70.

\textbf{Key Words:} Lieb-Thirring inequalities, Schr\"{o}dinger operator,
operators on groups, quantum graphs, Anderson model.
\end{abstract}

\section{Introduction}

Let us recall the classical estimates concerning the negative eigenvalues of the
operator $H=-\Delta +V(x)$ on $L^{2}(R^{d}),$ $d\geq 3.$ Let $N_{E}(V)$ be
the number of eigenvalues $E_{i}$ of the operator $H$ that are below or equal to $%
E\leq 0$. In particular, $N_{0}(V)$ is the number of non-positive
eigenvalues. Let
\[
N(V)=\#\{E_{i}<0\}
\]
be the number of strictly negative eigenvalues of the operator $H$. Then the
Cwikel-Lieb-Rozenblum and Lieb-Thirring inequalities have the following
form, respectively, (see \cite{c}, \cite{L}-\cite{Lt1}, \cite{r}, \cite{rs})
\begin{equation}
N(V)\leq C_{d}\int_{R^{d}}W^{\frac{d}{2}}(x)dx,  \label{c}
\end{equation}
\begin{equation}
\sum_{i:E_{i}<0}|E_{i}|^{\gamma }\leq C_{d,\gamma }\int_{R^{d}}W^{\frac{d}{2}%
+\gamma }(x)dx.  \label{l}
\end{equation}
Here $W=|V_{\_}|,~V_{\_}(x)=\min (V(x),0),$ $d\geq 3,~\gamma \geq 0.$ The
inequality (\ref{c}) can be considered as a particular case of (\ref{l})
with $\gamma =0$. Conversely, the inequality (\ref{l}) can be easily derived
from (\ref{c}) (see \cite{rs}). So, below we will mostly discuss the
Cwikel-Lieb-Rozenblum inequality and its extensions, although some new
results concerning the Lieb-Thirring inequality will also be stated.

A review of different approaches to the proof of (\ref{c}) can be found in \cite{1}.
We will remind only several results. E. Lieb \cite{L}, \cite{L1} and I. Daubechies \cite{d} offered the following general form of
(\ref{c}) and (\ref{l}).
Let $H=H_{0}+V(x)$, and $V(x)=V_{+}(x)-V_{-}(x),$ $V_{\pm }\geq 0.$\ Then
\begin{equation}
N(V)\leq \frac{1}{g(1)}\int_{0}^{\infty }\frac{\pi (t)}{t}%
dt\int_{X}G(tW(x))\mu (dx).  \label{rozs}
\end{equation}
\begin{equation}
\sum_{i:E_{i}<0}|E_{i}|^{\gamma }\leq \frac{1}{g(1)}\int_{0}^{\infty }\frac{\pi (t)}{t}%
dt\int_{X}G(tW(x))W^{\gamma}(x)\mu (dx).  \label{e2a}
\end{equation}
Here $W=V_{-}=\max (0,-V(x)),~G$ is a continuous, convex, non-negative
function which grows at infinity not faster than a polynomial, and is such
that $z^{-1}G(z)$ is integrable at zero (hence, $G(0)=0$), and the integral (%
\ref{rozs}) is finite. The function $g(\lambda ),$ $\lambda \geq 0,$\ is
defined by
\begin{equation}
g(\lambda )=\int_{0}^{\infty }z^{-1}G(z)e^{-z\lambda }dz,\text{ \ i.e. }
g(1)=\int_{0}^{\infty }z^{-1}G(z)e^{-z}dz.  \label{rozs1}
\end{equation}
Note that $\pi (t)=(2\pi t)^{-\frac{d}{2}}$ in the classical case of $%
H_{0}=-\Delta $ on $L^{2}(R^{d})$, and (\ref{c}) follows from (\ref{rozs})
in this case by substitution $t\rightarrow \tau =tW(x)$ if $G$ is such that $%
\int_{0}^{\infty }z^{-1-\frac{d}{2}}G(z)dz<\infty .$

The inequalities above are meaningful only for those $W$ for which integrals
converge. They become particularly transparent (see \cite{L1}) if $G(z)=0$\ for $z\leq
\sigma ,$ $G(z)=z-\sigma $\ for $z>\sigma ,$\ $\sigma \geq 0.$\ Then (\ref
{rozs}), (\ref{e2a}) take the form
\begin{eqnarray}
N(V) &\leq &\frac{1}{c(\sigma )}\int_{X}W(x)\int_{\frac{\sigma }{W(x)}%
}^{\infty }\pi (t)dt\mu (dx),  \label{1xx} \\
\sum_{i:E_{i}<0}|E_{i}|^{\gamma } &\leq &\frac{1}{c(\sigma )}%
\int_{X}W^{\gamma +1}(x)\int_{\frac{\sigma }{W(x)}}^{\infty }\pi (t)dt\mu (dx),
\label{1yy}
\end{eqnarray}
where $c(\sigma )=e^{-\sigma }\int_{0}^{\infty }\frac{ze^{-z}dz}{z+\sigma }.$

I. Daubichies \cite{d} used Lieb method to justify the estimates above for
some pseudo-differential operators in $R^{d}$. She also mentioned there that the
Lieb method works in a wider setting. A slightly different approach based
 on the
Trotter formula was used by G. Rozenblum and M. Solomyak \cite{2},
\cite{1}. They proved (\ref{rozs}) for a wide class of operators
in $L^2(X,\mu)$ where $X$ is a measure space with a $\sigma
$-finite\ measure $\mu=\mu(dx)$. They also suggested the following
form of (\ref{rozs}). Assume that the function $\pi (t)$ has
different power asymptotics as $t\rightarrow 0$ and $t\rightarrow
\infty $. Let
\begin{equation}
p_{0}(t,x,x)\leq c/t^{\alpha /2},~~~t\leq h,~~~p_{0}(t,x,x)\leq c/t^{\beta
/2},~~~t>h,  \label{pipi}
\end{equation}
where $h>0$ is arbitrary$.$ The parameters $\alpha $ and $\beta $
characterize the ``local dimension" and the ``global dimension" of $X$,
respectively. For example $\alpha =\beta =d$ in the classical case of the
Laplacian $H_{0}=-\Delta $ in the Euclidean space $X=R^{d}$. If $%
H_{0}=-\Delta $ is the difference Laplacian on the lattice $X=Z^{d}$, then $%
\alpha =0,$ $\beta =d.$ If $X=S^{n}\times R^{d}$ is the product of $n$%
-dimensional sphere and $R^{d}$, then $\alpha =n+d,$ $\beta =d$.

If $\alpha ,\beta >2,$ inequality (\ref{rozs}) implies (see \cite{1}) that
\begin{equation}
N(V)\leq C(h)[\int_{\{W(x)\leq h^{-1}\}}W^{\frac{\beta }{2}}(x)\mu
(dx)+\int_{\{W(x)>h^{-1}\}}W^{\frac{\alpha }{2}}(x)\mu (dx)],  \label{mr}
\end{equation}
Note that the restriction $\beta >2$ is essential here in the same way as the
condition $d>2$ in (\ref{c}). We will show that the assumption on $\alpha $
can be omitted, but the form of the estimate in (\ref{mr}) changes in this
case.

The paper consists of two parts. In a shorter first part we will give a detail proof of the
general form of Cwikel-Lieb-Rozenblum (\ref{rozs}) and Lieb-Thirring (\ref{e2a}) inequalities
for Schr\"{o}dihger operator in $L^2(X,\mu)$ where $X$ is a metric space with a $\sigma $-finite\ measure $\mu$.
We shall use the Lieb method which is based on trace inequalities and
the Kac-Feynman representation of the Schr\"{o}dinger parabolic semigroup. This approach could be particularly preferable for readers with a background in
probability theory. We do not go there beyond results
obtained in \cite{2}, \cite{1}. This part has mostly a methodological character. We also will show that inequality
(\ref{rozs}) is valid for $N_0(V)$, not only for $N(V)$.

The main goal of the paper is a new set of examples. We will consider
operators which may have different power asymptotics of $\pi (t)$ as $%
t\rightarrow 0$ or $t\rightarrow \infty $ or exponential asymptotics as $%
t\rightarrow \infty $. The latter case will allow us to consider the potentials
which decay very slowly at infinity. This is particularly important in some
applications, such as Anderson model, where the borderline between operators
with a finite and infinite number of eigenvalues is defined by the decay of
the perturbation in the logarithmic scale.

The paper is organized as follows. The general statement will be
proved in Theorem \ref{t1} in the next section. Theorems \ref{c1},
\ref{texp1} at the end of that section are consequences of Theorem
\ref{t1}. They provide more transparent results under additional
assumptions on the asymptotic (power or exponential) behavior of
$\pi (t).$ Note that we consider all $\alpha \geq 0$ in
(\ref{pipi}). Sections 3-6 are devoted to examples. Some cases of
a low local dimension $ \alpha $ are studied in Section 3.
Operators on lattices (see also (\cite{1})) and graphs are
considered there. Section 4 deals with perturbations of Anderson
operator. Lobachevsky plane (see also (\cite{1})) and pseudo
differential operators related to processes with independent
increments are considered in Section 5. Section 6 is devoted to
operators on free groups, continuous and discrete Heisenberg group
(see also (\cite{Gav}),(\cite{G})), continuous and discrete groups
of affine transformations of the line. The Appendix contains the
justification of the asymptotics of $\pi (t)$ for the quantum
graph operator.

Note that in order to apply any of estimates (\ref{rozs}),(\ref{e2a}) or (\ref{1xx})-(\ref{mr})
 one needs an exact bound for $\pi(t)$ which can be a challenging problem in some cases.

The authors are very grateful to V. Konakov and O. Safronov for very useful
discussions.

\section{General Cwikel-Lieb-Rozenblum and\newline
Lieb-Thirring inequalities.}

We will assume that $X$ is a complete $\sigma $-compact metric space with
Borel $\sigma $-algebra $\mathcal{B}(X)$ and a $\sigma $-finite measure $\mu
(dx).$ Let $H_{0}$ be a self-adjoint non-negative operator on $L^{2}(X,%
\mathcal{B,}\mu )$ with the following two properties:

(a) Operator $-H_0$ is the generator of a semigroup $P_t $ acting
on $C(X)$. The kernel $p_0(t,x,y)$ of $P_t$ is continuous with
respect to all the variables when $t>0$ and satisfies the relations
\begin{equation}
\frac{\partial p_{0}}{\partial t}=-H_{0}p_{0},~~~t>0,~~~p_{0}(0,x,y)=\delta
_{y}(x),~~\int_{X}p_{0}(t,x,y)\mu (dy)=1,  \label{para}
\end{equation}
i.e. $p_0$ is a fundamental solution
of the corresponding parabolic problem.
We assume that $p_{0}(t,x,y)$ is symmetric, non-negative, and it defines a Markov process $%
x_{s},$ $s\geq 0,$\ on $X$ with the transition density $p_{0}(t,x,y)$ with
respect to the measure $\mu $.

Note that this assumption implies that $p_{0}(t,x,x)$ is strictly positive
for all $x\in X,$ $t>0$, since
\begin{equation}
p_{0}(t,x,x)=\int_{X}p_{0}^2(\frac{t}{2},x,y)\mu (dy)>0.  \label{p00}
\end{equation}

(b) There exists a function $\pi (t)$ such that $p_{0}(t,x,x)\leq \pi (t)$
for $t\geq 0$ and all $x\in X.$ We also assume that $\pi (t)$ has at most
power singularity at $t\rightarrow 0$ and is integrable at infinity, i.e.
there exists $m$ such that
\begin{equation}
\int_{0}^{\infty }\frac{t^{m}}{1+t^{m}}\pi (t)dt<\infty .  \label{conb}
\end{equation}
Note that condition (b) implies that
\begin{equation}
p_{0}(t,x,y)\leq \pi (t),~~x,y\in X.  \label{conb1}
\end{equation}
In fact,
\[
p_{0}(t,x,y)=\int_{X}p_{0}(\frac{t}{2},x,z)p_{0}(\frac{t}{2},z,y)\mu
(dz)\leq (\int_{X}p_{0}^{2}(\frac{t}{2},x,z)\mu (dz))^{\frac{1}{2}%
}(\int_{X}p_{0}^{2}(\frac{t}{2},z,y)\mu (dz))^{\frac{1}{2}},
\]
which implies (\ref{conb1}) due to (\ref{p00}). Let us note that (\ref{conb}%
), (\ref{conb1}) imply that the process $x_{s}$ is transient.

We decided to put an extra requirement on $X$ to be a metric space
in order to be able to assume that $p_{0}$ is continuous and use a
standard version of the Kac-Feynman formula. This makes all the
arguments more transparent. In fact, $X$ is a metric space in all
examples below. However, all the arguments can be modified to be
applicable to the case when $X$ is a measure space by using
$L^{2}$-theory of Markov processes based on the Dirichlet forms.

Many examples of operators which satisfy conditions (a) and (b) will be
given later. At this point we would like to mention only a couple of
examples. First, note that self-adjoint uniformly elliptic operators of second order satisfy
conditions (a) and (b). Condition (b) holds with $\pi (t)=Ct^{-d/2}$ due to
Aronson inequality.

Another wide class of operators with conditions (a) and (b) consists of
operators which satisfy condition (a) and are invariant with respect to
transformations from a rich enough subgroup $\Gamma $ of the group of
isometries of $X.$ The subgroup $\Gamma $ has to be transitive, i.e., for
some reference point $x_{0}\in X$ and each $x\in X$ there exists an element $%
g_{x}\in \Gamma $ for which\ $g_{x}(x_{0})=x.$ \ Then $%
p_{0}(t,x,x)=p_{0}(t,x_{0},x_{0})=\pi (t).$ The simplest example of such an
operator is given by $H_{0}=-\Delta $ on \ $L^{2}(R^{d},\mathcal{B}(R^{d})%
\mathcal{,}dx).$ The group $\Gamma $ in this case is the group of
translations or the group of all Euclidean transformations (translations and
rotations). Another example is given by $X=Z^{d}$ being a lattice and $%
-H_{0} $ a difference Laplacian. Other examples will be given later.

(c) Our next assumption mostly concerns the potential. We need to know that
the perturbed operator $H=H_{0}+V(x)$ is well defined and has pure discrete
spectrum on the negative semiaxis. For this purpose it is enough to assume
that the operator $V(x)(H_{0}-E)^{-1}$ is compact for some $E>0$. This
assumption can be weakened. If the domain of $H_{0}$ contains a dense in $%
L^{2}(X,\mathcal{B,}\mu )$ set of bounded compactly supported functions,
then it is enough to assume that $V_{-}(x)(H_{0}-E)^{-1}$\ is compact for
some $E>0$ and the positive part of the potential is locally integrable (see
\cite{bs}).

%For example,
%if $n=2$, this condition is equivalent to
%\[
%\int_{X}\int_{X}W(x)W(y)R^2_E(x,y)d\mu(x)d\mu(y)< \infty.
%\]
%In the case of the standard Schrodinger operator in $R^3$ and
%$E=0$ the latter means that
%\[
%\int_{R^3}\int_{R^3}\frac{W(x)W(y)}{|x-y|^2}dxdy< \infty,
%\]
%i.e. $W$ belongs to the Rollnik class.

Typically (in particular, in all the examples below) $H_{0}$ is an elliptic
operator, the kernel of the resolvent $(H_{0}-E)^{-1}$ has singularity only
at $x=y$, this singularity is weak, and the assumptions (c) holds if the
potential has an appropriate behavior at infinity. %at least for
%$W\in C_{com}(X)$. Thus condition (c) requires only some type of
%decay of $W$ at infinity and restricts the type of singularities
%that $W$ may have.
Therefore we do not need to discuss the validity of this assumption in the
examples below.

\begin{theorem}
\label{t1}Let $(X,\mathcal{B,}\mu )$ be a complete $\sigma $-compact metric
space with the Borel $\sigma $-algebra $\mathcal{B}$ and a $\sigma $-finite\
measure $\mu $\ on $\mathcal{B}.$

Let $H=H_{0}+V(x)$, where $H_{0}$ is a self-adjoint, non-negative operator
on $L^{2}(X,\mathcal{B,}\mu )$, the potential $V=V(x)=V_{+}-V_{-},$ $V_{\pm
}\geq 0,$\ is real valued, and the assumptions (a)-(c) hold.

Then
\begin{equation}
N_{0}(V)\leq \frac{1}{g(1)}\int_{0}^{\infty }\frac{\pi (t)}{t}%
\int_{X}G(tW(x))\mu (dx)dt,  \label{e1}
\end{equation}
and
\begin{equation}
\sum_{i:E_{i}<0}|E_{i}|^{\gamma }\leq \frac{1}{g(1)}\int_{0}^{\infty }\frac{%
\pi (t)}{t}\int_{X}G(tW(x))W(x)^{\gamma }\mu (dx)dt,  \label{e2}
\end{equation}
where $W(x)=V_{-}(x),$ and functions $G$ and$\ g$ are introduced above in (%
\ref{rozs}) and (\ref{rozs1}).

%The integrals in the right-hand sides above converge if
%$G$ is linear at infinity and $G<Cz^{m+1},~ z\rightarrow0,$ where $m$ is defined in the condition (c).
\end{theorem}

\begin{remark}
Note that (\ref{e1}) differs from (\ref{rozs}) only by inclusion of the
dimension of the null space of the operator $H$ into the left-hand side of (%
\ref{e1}). This difference is not very essential, and the first goal of this
part of the paper is to give an alternative proof of (\ref{rozs}) suitable
for readers with a background in probability theory.
\end{remark}

\begin{remark}
\label{r2}If $G(z)=0$\ for $%
z\leq \sigma ,$ $G(z)=z-\sigma $\ for $z>\sigma ,$\ $\sigma \geq 0$, then (%
\ref{e1}), (\ref{e2}) take the form
\begin{eqnarray}
N_{0}(V) &\leq &\frac{1}{c(\sigma )}\int_{X}W(x)\int_{\frac{\sigma }{W(x)}%
}^{\infty }\pi (t)dt\mu (dx),  \label{xx} \\
\sum_{i:E_{i}<0}|E_{i}|^{\gamma } &\leq &\frac{1}{c(\sigma )}%
\int_{X}W^{\gamma +1}(x)\int_{\frac{\sigma }{W(x)}}^{\infty }\pi (t)dt\mu (dx),
\label{yy}
\end{eqnarray}
where $c(\sigma )=e^{-\sigma }\int_{0}^{\infty }\frac{ze^{-z}dz}{z+\sigma }.$
Some applications of these inequalities will be given below.
\end{remark}

\begin{remark}
\label{r}Inequalities (\ref{e1}), (\ref{e2}) are valid with $\pi
(t)$ moved under sign of the interior integrals and replaced by
$p_{0}(t,x,x).$ For example, (\ref{e1}) holds in the following
form
\[
N_{0}(V)\leq \frac{1}{g(1)}\int_{0}^{\infty }\frac{1}{t}%
\int_{X}p_0(t,x,x)G(tW(x))\mu (dx)dt.
\]
The same change can be made in (\ref{xx}), (\ref{yy}). A very
minor change in the proof of the theorem is needed in order to
justify this remark. Namely, one needs only to omit the last line
in (\ref{last5}).
\end{remark}
\textbf{Proof of Theorem \ref{t1}.} \textit{Step 1}. Since the eigenvalues $%
E_{i}$ depend monotonically on the potential $V(x),$ without loss of
generality one can assume that $V(x)=-W(x)\leq 0.$

First (steps 1-6), we'll prove inequality (\ref{e1}) for $N(V)$ instead of $%
N_0(V)$. Here we can assume that $V(x)\in C_{\mathrm{com}}(X)$. Indeed,
when $N(V)$ is considered, inequality (\ref{e1}) with $V(x)\in C_{\mathrm{com%
}}(X)$ implies the same inequality with any $V$ such that the
integral in (\ref{e1}) converges (see \cite{rs}). Then (step 7), we'll show
that inequality (\ref{e1}) for $N(V)$ leads to the same inequality for $%
N_{0}(V)$. Finally (step 8), we will remind the reader of standard arguments
which allow us to derive (\ref{e2}) from (\ref{e1}).

\textit{Step 2}. We denote by $B$ and $B_{n}$ the operators
\[
B=W^{1/2}(H_0+\varkappa^2)^{-1}W^{1/2},~~B_{n}=W^{1/2}(H_0+\kappa^2+n
W)^{-1}W^{1/2},~~W=W(x).
\]
If $N_{-\varkappa^2}(V)=\#\{E_{i}\leq-\varkappa ^{2}<0\}$, $\lambda_k$ are
eigenvalues of the operator $B$ and $n(\lambda, B)=\#\{k:\lambda_k \geq
\lambda \}$, then the Birman-Schwinger principle implies
\begin{equation}
N_{-\varkappa ^{2}}(V) = n(1,B).
\end{equation}
Thus, if $F=F(\lambda),~\lambda \geq 0, $ is a non-negative strictly
monotonically growing function, and $\{\mu_k\}$ is the set of eigenvalues of
the operator $F(B)$, then
\begin{equation}
N_{-\varkappa ^{2}}(V) \leq \sum _{k:\mu_k \geq F(1)}1 \leq \frac
{1}{F(1)}\sum _{k:\mu_k \geq F(1)}\mu _k \leq \frac {1}{F(1)} \mathrm{{Tr}%
F(B).}  \label{bs1}
\end{equation}
This inequality will be used with the function $F$ of the form
\begin{equation}
F(\lambda)=\int_0^{\infty}P(e^{-z})e^{\frac{-z}{ \lambda}}dz,~~P(t)=
\sum_0^Nc_nt^n,  \label{nf}
\end{equation}
The exponential polynomial $P(e^{-z}),~z>0,$ will be chosen later, but it
will be a non-negative function with zero of order $m$ at $z=0$, i.e.
\begin{equation}
P(e^{-z}) \leq C\frac{z^{m}}{1+z^{m}},~~z\geq0,  \label{pat0}
\end{equation}
where $m$ is defined in the condition (b). Since $P(e^{-z})\geq 0$, (\ref{nf}%
) implies that $F$ is non-negative and monotonic, and therefore (\ref{bs1})
holds.

From (\ref{nf}) it follows that
\[
F(\lambda )=\sum_{n=0}^{N}c_{n}\frac{\lambda }{1+n\lambda },
\]
and the obvious relation $B_{n}=B(1+nB)^{-1}$ implies that
\[
F(B)=\sum_{n=0}^{N}c_{n}B_{n}=W^{\frac{1}{2}}
\sum_{n=0}^{N}c_{n}(H_{0}+\kappa ^{2}+nW)^{-1}W^{\frac{1}{2}}.
\]

For an arbitrary operator $K$, we denote its kernel by $K(x,y)$. The kernel
of the operator $F(B)$ can be expressed trough the fundamental solutions $%
p=p_{n}(t,x,y)$ of the parabolic problem
\[
p_{t}=(H_{0}+nW(x))p,~t>0,~~p(0,x,y)=\delta _{y}(x).
\]
Namely,
\begin{equation}
F(B)(x,y)=W^{\frac{1}{2}}(x)\int_{0}^{\infty }e^{-\kappa
^{2}t}\sum_{n=0}^{N}c_{n}p_{n}(t,x,y)dtW^{\frac{1}{2}}(y).  \label{unc}
\end{equation}
It will be shown below that the integral above converges uniformly in $x$
and $y$ when $\kappa =0$. Hence, the kernel $F(B)(x,y)$ is continuous. Since
the operator $F(B)$ is non-negative, from the last relation and (\ref{bs1}),
after passing to the limit as $\kappa \rightarrow 0$, it follows that
\begin{equation}
N(V)\leq \frac{1}{F(1)}\int_{0}^{\infty
}\int_{X}W(x)\sum_{n=0}^{N}c_{n}p_{n}(t,x,x)dt\mu (dx).  \label{16}
\end{equation}

\textit{Step 3}. The Kac-Feynman formula allows us to write an ''explicit''
representation for the Schr\"{o}dinger semigroup $e^{t(-H_{0}-nW(x))}$ using
the Markov process $x_{s}$ associated to the unperturbed operator $H_{0}.$
Namely, the solution of the parabolic problem
\begin{equation}
\frac{\partial u}{\partial t}=-H_{0}u-nW(x)u,~~~t>0,~~~u(0,x)=\varphi (x)\in
C(X),  \label{par}
\end{equation}
can be written in the form
\[
u(t,x)=E_{x}e^{-n\int_{0}^{t}W(x_{s})ds}\varphi (x_{t}).
\]

Note that the finite-dimensional distributions of $x_{s}$ (for $%
0<t_{1}<...<t_{n},$ \ $\Gamma _{1},...\Gamma _{n}\in \mathcal{B}(X)$ ) are
given by the formula
\begin{eqnarray*}
P_{x}(x_{t_{1}}\in\Gamma _{1},...,x_{t_{n}}\in \Gamma _{n})
\end{eqnarray*}
\begin{eqnarray*}
=\int_{\Gamma _{1}}...\int_{\Gamma
_{n}}p_{0}(t_{1},x,x_{1})p_{0}(t_{2}-t_{1},x_{1},x_{2})...p_{0}(t_{n}-t_{n-1},x_{n-1},x_{n})\mu (dx_{1})...\mu (dx_{n}).
\end{eqnarray*}
If $p_{0}(t,x,y)>0$, then one can define the conditional process (bridge) $%
\widehat{b}_{s}=\widehat{b}_{s}^{x\rightarrow y,t},$ $s\in \lbrack 0,t],$
which starts at $x$ and ends at $y.$ Its finite-dimensional distributions
are
\begin{eqnarray*}
P_{x\rightarrow y}(\widehat{b}_{t_{1}}\in\Gamma _{1},...,\widehat{b}%
_{t_{n}}\in \Gamma _{n})
\end{eqnarray*}
\begin{eqnarray*}
=\frac{\int_{\Gamma _{1}}...\int_{\Gamma
_{n}}p_{0}(t_{1},x,x_{1})...p_{0}(t_{n}-t_{n-1},x_{n-1},x_{n})p_{0}(t-t_{n},x_{n},y)\mu (dx_{1})...\mu (dx_{n})%
}{p_{0}(t,x,y)}.
\end{eqnarray*}
In particular, the bridge $\widehat{b}_{s}^{x\rightarrow x,t},$ $s\in
\lbrack 0,t],$ is defined, since $p_{0}(t,x,x)>0$ (see condition (a)).

Let $p=p_n(t,x,y)$ be the fundamental solution of the problem (\ref{par}).
Then $p_n(t,x,y)$ can be expressed in terms of the bridge $\widehat{b}_{s}=%
\widehat{b}_{s}^{x\rightarrow y,t},$ $s\in \lbrack 0,t]:$
\begin{equation}
p_n(t,x,y)=p_{0}(t,x,y)E_{x\rightarrow y}e^{-n\int_{0}^{t}W(\widehat{b}%
_{s})ds}.  \label{kf}
\end{equation}

One of the consequences of (\ref{kf}) is that
\begin{equation}
p_n(t,x,y)\leq p_{0}(t,x,y).  \label{mp}
\end{equation}

Another consequence of (\ref{kf}) is the uniform convergence of the integral
in (\ref{unc}) (and in (\ref{16})). In fact, (\ref{pat0}) implies that
\[
\sum_{n=0}^Nc_ne^{-n\int_0^tW(\widehat{b}_{s})ds}\leq C\frac{t^m}{1+t^m}.
\]
Hence from (\ref{kf}) and (\ref{conb1}) it follows that the integrand in (%
\ref{unc}) can be estimated from above by $C\pi(t)\frac{t^m}{1+t^m}$. Then
the uniform convergence of the integral in (\ref{unc}) follows from (\ref
{conb}).

Now (\ref{16}) and (\ref{kf}) imply
\[
N(V)\leq \frac {1}{F(1)}\int_{0}^{\infty
}\int_{X}W(x)p_{0}(t,x,x)E_{x\rightarrow
x}[\sum_{n=0}^Nc_ne^{-n\int_{0}^{t}W(\widehat{b}_{s})ds}]\mu (dx)dt,~~\widehat{%
b}_{s}=\widehat{b}_{s}^{x\rightarrow x,t}.
\]

\textit{Step 4}. We would like to rewrite the last inequality in the form
\begin{equation}
N(V)\leq \frac {1}{F(1)}\int_{0}^{\infty
}\int_{X}p_{0}(t,x,x)E_{x\rightarrow x}[W(\widehat{b}_{\tau
})\sum_{n=0}^Nc_ne^{-n\int_{0}^{t}W(\widehat{b}_{s})ds}]\mu (dx)dt  \label{ins}
\end{equation}
with an arbitrary $\tau \in \lbrack 0,t].$ For that purpose, it is enough to
show that
\begin{eqnarray}
&&\int_{X}p_{0}(t,x,x)E_{x\rightarrow x}[W(\widehat{b}_{\tau
})e^{-\int_{0}^{t}mW(\widehat{b}_{s})ds}]\mu (dx)  \nonumber \\
&=&\int_{X}p_{0}(t,x,x)W(x)E_{x\rightarrow x}[e^{-\int_{0}^{t}mW(\widehat{b}%
_{s})ds}]\mu (dx).  \label{ab}
\end{eqnarray}
The validity of (\ref{ab}) can be justified using the Markov property of $%
\widehat{b}_{s}$ and its symmetry (reversibility in time). We fix $\tau \in
(0,t).$ Let $y=\widehat{b}_{\tau }.$ We split $\widehat{b}_{s}$ into two
bridges $\widehat{b}_{u}^{x\rightarrow y,\tau },$ $u\in \lbrack 0,\tau ],$
and $\widehat{b}_{v}^{y\rightarrow x,t},$ $v\in \lbrack \tau ,t].$ The first
bridge starts at $x$ and ends at $y$, the second one starts at $y$ and goes
back to $x$. Using these bridges, one can represent the left hand side above
as
\begin{eqnarray*}
&&\int_{X}\int_{X}W(y)[p_{0}(\tau ,x,y)p_{0}(t-\tau ,y,x)-p_m(\tau
,x,y)p_m(t-\tau ,y,x)]\mu (dx)\mu (dy) \\
&=&\int_{X}W(y)[p_{0}(t,y,y)-p_m(t,y,y)]\mu (dy),
\end{eqnarray*}
which coincides with the right hand side of (\ref{ab}). This proves (\ref
{ins}).

\textit{Step 5. }We take the average of both sides of (\ref{ins}) with
respect to $\tau \in \lbrack 0,t]$ and rewrite it in the form
\begin{eqnarray}
N(V) &\leq &\frac{1}{F(1)}\int_{0}^{\infty }\int_{X}\frac{p_{0}(t,x,x)}{t}%
E_{x\rightarrow x}\sum_0^N(c_m\int_{0}^{t}W(\widehat{b}_{s})dse^{-%
\int_{0}^{t}mW( \widehat{b}_{s})ds})\mu (dx)dt  \nonumber \\
&=&\frac{1}{F(1)}\int_{0}^{\infty }\int_{X}\frac{p_{0}(t,x,x)}{t}%
E_{x\rightarrow x}(uP(e^{-u}))\mu (dx)dt,~~~u=\int_{0}^{t}W(\widehat{b}_{s})ds,
\label{ab1}
\end{eqnarray}
where $P$ is the polynomial defined in (\ref{nf}) and (\ref{16}).

Let now $P$ be such that
\begin{equation}
uP(e^{-u})\leq G(u),  \label{pg}
\end{equation}
where $G$ is defined in the statement of Theorem \ref{t1}. Then one can
replace $uP(e^{-u})$ in (\ref{ab1}) by $G(u)$. Then the Jensen inequality
implies that
\[
G (\int_{0}^{t}W(\widehat{b}_{s}))ds=G (\frac{1}{t}\int_{0}^{t}tW(\widehat{b}%
_{s}))ds\leq \frac{1}{t}\int_{0}^{t}G (tW(\widehat{b}_{s}))ds.
\]
This allows us to rewrite (\ref{ab1}) in the form
\begin{equation}
N(V)\leq \frac{1}{F(1)}\int_{0}^{\infty }\int_{X}\frac{p_{0}(t,x,x)}{t}%
\frac{1}{t}\int_{0}^{t}E_{x\rightarrow x}G (tW(\widehat{b}_{s}))ds\mu (dx)dt.
\label{nge}
\end{equation}
It is essential that one can use the exact formula for the distribution
above:
\[
E_{x\rightarrow x}G (tW(\widehat{b}_{s}))=\int_{X}G (tW(z))\frac{%
p_{0}(s,x,z)p_{0}(t-s,z,x)}{p_{0}(t,x,x)}\mu (dz).
\]
From here and (\ref{nge}) it follows that
\begin{eqnarray}
N(V) &\leq & \frac{1}{F(1)}\int_{0}^{\infty }\frac{1}{t^{2}}%
\int_{0}^{t}ds\int_{X} \int_{X}G(tW(z))p_{0}(s,x,z)p_{0}(t-s,z,x)\mu
(dx)\mu (dz)dt  \nonumber \\
&=&\frac{1}{F(1)}\int_{0}^{\infty }\frac{1}{t^{2}}\int_{0}^{t}ds\int_{X}\mu
(dz)G (tW(z))p_{0}(t,z,z)dt  \nonumber \\
&=&\frac{1}{F(1)}\int_{0}^{\infty }\frac{1}{t}\int_{X}G
(tW(z))p_{0}(t,z,z)\mu (dz)dt  \nonumber \\
&\leq &\frac{1}{F(1)}\int_{0}^{\infty }\frac{\pi(t)}{t}\int_{X}G
(tW(z))\mu (dz)dt,  \label{last5}
\end{eqnarray}
where $F(1)$ is defined in (\ref{nf}).

\textit{Step 6}. Now we are going to specify the choice of the polynomial $P$
which was used in the previous steps. It must be non-negative and satisfy (%
\ref{conb}) and (\ref{pg}). Polynomial $P$ will be determined by the choice
of the function $G$. Note that it is enough to prove (\ref{e1}) for
functions $G$ which are linear at infinity. In fact, for arbitrary $G$, let $%
G_N\leq G$ be a continuous function which coincides with $G$ when $z \leq N$
and is linear when $z \geq N$. For example, if $G$ is smooth, $G_N$ can be
obtained if the graph of $G$ for $z \geq N$ is replaced by the tangent line
through the point $(N, G(N)$. Since $G_N \leq G$, the validity of (\ref{e1})
for $G_N$ implies (\ref{e1}) with the function $G$ in the integrand and $%
g(1) $ being replaced by $g_N(1)$. Passing to the limit as $N\rightarrow
\infty$ in this inequality, one gets (\ref{e1}), since $g_N(1)\rightarrow
g(1) $ as $N\rightarrow \infty$. Similar arguments allow us to assume that $%
G =0$ in a neighborhood of the origin (The validity of (\ref{e1}) for $%
G_{\varepsilon}(z)=G(z-\varepsilon) \leq G(z)$ implies (\ref{e1})). Now
consider $G^{\varepsilon}(z)=$ max$(G(z),y(\varepsilon,z))$ where $%
y(\varepsilon,z))=z^{m+1}, z\leq
\varepsilon,~y(\varepsilon,z)=(m+1)(z-\varepsilon)+\varepsilon ^{m+1}, z>
\varepsilon$, with $m$ defined in condition (b). We will show later that the
right-hand side of (\ref{e1}) is finite for $G=G^{\varepsilon}$. Thus if (%
\ref{e1}) is proved for $G=G^{\varepsilon}$, then passing to the limit as $%
\varepsilon \rightarrow0 $ one gets (\ref{e1}) for $G$. Hence we can assume
that $G=az$ at infinity and $G=z^{m+1}$ in a neighborhood of the origin.
Note that $a\neq 0$, since $G$ is convex.

A special approximation of the function $G$ by exponential polynomials will
be used. Consider function $H(z)=\frac {G(z)}{z(1-e^{-z})^{m}}, z>0$. It is
continuous, nonnegative and has positive limits as $z\rightarrow 0$ and $%
z\rightarrow \infty$. Hence there is an exponential polynomial $%
p_{\varepsilon}(e^{-z})$ which approximates $H(z)$ from below, i.e.
\[
|H(z)-p_{\varepsilon}(e^{-z})|<\varepsilon,~~~0<p_{\varepsilon}(e^{-z})\leq
H(z)\leq 2p_{\varepsilon}(e^{-z}), ~~ z>0.
\]
In order to find $p_{\varepsilon}$, one can change the variable $t=e^{-z}$
and reduce the problem to the standard Weierstrass theorem on the interval
(0,1). If $P_{\varepsilon}(e^{-z})=(1-e^{-z})^{m}p_{\varepsilon}(e^{-z})$,
then
\begin{equation}  \label{pe1}
|z^{-1}G(z)-P_{\varepsilon}(e^{-z})|<\varepsilon,~0<P_{\varepsilon}(e^{-z})%
\leq z^{-1}G(z), ~~ z>0;~~~ P_{\varepsilon}(e^{-z})<Cz^{m},~z\rightarrow 0.
\end{equation}
%and
%\begin{equation}
%z^{-1}G(z)\leq 2P_{\varepsilon}(e^{-z}),~~z>0. \label{pe2}
%\end{equation}

We will choose polynomial $P$ in (\ref{nf}) and (\ref{16}) to be equal to $%
P_{\varepsilon}$. The last two of relations (\ref{pe1}) show that $%
P=P_\varepsilon$ satisfies all the properties used to obtain (\ref{last5}).
Function $F$ in (\ref{last5}) is defined by (\ref{nf}) with $P=P_\varepsilon$%
, and therefore $F(1)=F_{\varepsilon}(1)$ depends on $\varepsilon$. From the
first relation of (\ref{pe1}) it follows that $F_{\varepsilon}(1)
\rightarrow g(1)$ as $\varepsilon \rightarrow 0$. Thus passing to the limit
in (\ref{last5}) as $\varepsilon \rightarrow 0$ we complete the proof of
inequality (\ref{e1}) for $N(V)$.

%It was shown above that the integral in (\ref{e1}) converges if function $G(u)$ is replaced by $uP(e^{-u})$ where non-negative %exponential polynomial $P(e^{-u})$ has zero of order $m$ at the origin. Thus (\ref{pe2}) implies convergence of this integral for %convex non-negative functions $G$ which are linear at infinity and have zero of order $m+1$ at the origin.

\textit{Step 7}. Now we are going to show that inequality (\ref{e1}) for $%
N(V)$ implies the validity of this inequality for $N_{0}(V)$ under the
assumption that integral (\ref{e1}) converges. We can assume that $G$ is
linear at infinity and $G(z)=z^{m+1}$ in a neighborhood of the origin (see
step 6). Then $G(2tW(x))\leq C G(tW(x))$, and therefore the convergence
of the integral (\ref{e1}) implies the convergence of the same integral with $W$ replaced by $2W$.

Let $ n$ be the dimension of the null space of the
operator $H$. We need to show that $n$ is finite and $N(V)+n$ does not
exceed the right-hand side of (\ref{e1}).

Consider the operator
\[
H_{\varepsilon }=H+\varepsilon V(x)=H_{0}+(1+\varepsilon )V(x),~\varepsilon>0.
\]
The Dirichlet form of this operator
\[
(H_{\varepsilon }\phi,\phi)=(H\phi,\phi)+\varepsilon \int_{X}V(x)|\phi(x)|^{2}\mu (dx)
\]
is strictly negative on the space $T\backslash \{0\}$, where the $(N(V)+n)$-dimensional space $T$ is spanned by the eigenfunctions of $H$ with negative or zero eigenvalues.\footnote{This element of the proof in the previous versions of the paper was slightly inaccurate: the Dirichlet form of operator $H_\varepsilon$ was shown to be negative only on some basis in $T$, not on $T\backslash \{0\}$.  The authors are grateful to G. Rozenblioum who attracted their attention to this inaccuracy.}
Indeed, both terms on the right in the formula above are non positive on $T$. If $\phi\in T$ does not belong to the null space $N$ of $H$, then the first term is strictly negative. If $\phi\in N\backslash \{0\}$, then the second term is strictly negative since otherwise there exists $\phi=\phi_0\in N\backslash \{0\}$ such that $V\phi_0=0$. Then $\phi_0$ belongs to the null space of the unperturbed operator $H_0$. This contradicts the assumption (b) on the decay (integrability) of the heat kernel $p_0(t,x,x)$ as $t\to\infty$ (since $p_0\geq|\phi_0 (x)|^2$).

The negativity of the Dirichlet form on $T\backslash \{0\}$ implies that operator $H$ has
at least $N(V)+n$ strictly negative eigenvalues. Hence from inequality (\ref{e1}) for strictly
negative eigenvalues of the operator $H_{\varepsilon }$ it follows that
\begin{equation}
N(V)+n\leq \frac{1}{g(1)}\int_{0}^{\infty }\frac{\pi (t)}{t}%
\int_{X}G(t(1+\varepsilon)W (x))\mu (dx)dt.  \label{N+k}
\end{equation}

One may assume that the double integral in (\ref{e1}) converges. It was shown above that this assumption leads to the convergence of the integral in (\ref{N+k}) when $\varepsilon
=1 $. Then one can pass to the limit as $\varepsilon \rightarrow 0$ in (\ref
{N+k}) and get
\[
N(V)+n\leq \frac {1}{g(1)}\int_{0}^{\infty }\frac{\pi (t)}{t}%
\int_{X}G(tW(x))\mu (dx)dt.
\]
Hence (\ref{e1}) is
proved.

\textit{Step 8}. In order to prove (\ref{e2}), we note that
\[
\sum_{i:E_{i}<0}|E_{i}|^{\gamma }=\gamma \int_{0}^{\infty }E^{\gamma
-1}N_{E}(V)dE\leq \gamma \int_{0}^{\infty }E^{\gamma -1}N_{0}(-(W-E)_{+})dE
\]
\[
\leq \frac{\gamma }{g(1)}\int_{0}^{\infty }E^{\gamma -1}\int_{0}^{\infty }%
\frac{\pi (t)}{t}\int_{X}G(t(W(x)-E)_{+})\mu (dx)dtdE
\]
\[
=\frac{\gamma }{g(1)}\int_{0}^{\infty }\frac{\pi (t)}{t}\int_{X}%
\int_{0}^{W}E^{\gamma -1}G(t(W(x)-E))dE\mu (dx)dt
\]
\[
=\frac{\gamma }{g(1)}\int_{0}^{\infty }\frac{\pi (t)}{t}\int_{X}%
\int_{0}^{1}u^{\gamma -1}W^{\gamma }(x)G(tW(x)(1-u))du\mu (dx)dt.
\]
One can replace $G(tW(x)(1-u))$ here by $G(tW(x)),$ since $G$ is
monotonically increasing. This immediately implies (\ref{e2}).
%In order to complete the proof of the theorem it remains only to show that the integrals in right-hand
%side above are finite. The latter follows from the convergence of the integral (\ref{e1}), since there is a subset of $X$
%of positive measure where $W(x) <C < \infty$ and there is a subset of $X$
%of positive measure where $W(x) > \varepsilon >0$.
\qed

\begin{theorem}
\label{c1} %Let \[
%h_{m,\gamma }(v)=v^{\gamma }\int_{0}^{\infty }\pi (\frac{\tau }{v}%
%)(1-e^{-\tau })^{m}d \tau.
%\]
%Then
%\begin{equation}
%N_0(V) \leq C_m \int_{X}h_{m,0
%}(W(x))\mu (dx),~~~
%\sum_{i:E_{i}< 0}|E_{i}|^{\gamma }\leq C_{m,\gamma }\int_{X}h_{m,\gamma
%}(W(x))\mu (dx).  \label{c11}
%\end{equation}
Let $H=H_{0}+V(x)$, where $H_{0}$ is a self-adjoint, non-negative operator
on $L^{2}(X,\mathcal{B,}\mu )$, the potential $V=V(x)$ is real valued, and
the assumptions (a)-(c) hold.

If
\begin{equation}
\pi (t)\leq c/t^{\beta /2},\ \ t\rightarrow \infty ;\ \ \ \pi (t)\leq
c/t^{\alpha /2},\ \ t\rightarrow0  \label{c2}
\end{equation}
for some $\beta >2$ and $\alpha \geq 0$, then
\begin{equation}  \label{max}
N_0(V)\leq C(h)[\int_{X_h^-}W(x)^{\beta /2}\mu (dx)+\int_{X_h^+}bW(x)^{\max
(\alpha /2,1)}\mu (dx)],
\end{equation}
%\begin{equation}
%\sum_{i:E_{i}< 0}|E_{i}|^{\gamma }\leq C(h,\gamma
%)[\int_{X_h^-}W(x)^{\beta /2+\gamma }\mu
%(dx)+\int_{X_h^+}bW(x)^{\max (\alpha /2,1)+\gamma }\mu (dx)],
%\end{equation}
where $X_h^-=\{x:W(x)\leq h^{-1}\},~~X_h^+=\{x:W(x)> h^{-1}\},~~ b=1$ if $%
\alpha \neq 2,$ $b=ln(1+W(x))$ if $\alpha =2.$
\end{theorem}

In some cases $\max (\alpha /2,1)$ can be replaced by $\alpha /2$,
as will be discussed in Section 3.

\textbf{Proof of Theorem \ref{c1}}. We write (\ref{xx}) in the form $%
N_{0}(V)\leq I_{-}+I_{+},$ where $I_{\mp }$ correspond to integration in (%
\ref{xx}) over $X_{h}^{\mp }$, respectively.

Let $x\in X_{h}^{-}$, i.e., $W<h^{-1}.$ Then the interior integral in (\ref
{xx}) does not exceed
\begin{equation}
C(h)\int_{\frac{\sigma }{W}}^{\infty }t^{-\beta /2}dt=C(h)W^{(\beta /2)-1}.
\label{pow}
\end{equation}
Thus $I_{-}$ can be estimated by the first term in the right-hand side of (%
\ref{max}). Similarly,
\[
I_{+}\leq C(h)\int_{X_{h}^{+}}W(\int_{\frac{\sigma }{W}}^{h}+\int_{h}^{%
\infty })\pi (t)dt\leq C(h)\int_{X_{h}^{+}}W(\int_{\frac{\sigma }{W}%
}^{h}t^{-\alpha /2}dt+\int_{h}^{\infty }t^{-\beta /2}dt)dx,
\]
which does not exceed the second term in the right-hand side of (\ref{max}).
\qed

\begin{theorem}
\label{texp1} Let $H=H_{0}+V(x)$, where $H_{0}$ is a self-adjoint,
non-negative operator on $L^{2}(X,\mathcal{B,}\mu )$, the potential $V=V(x)$
is real valued, and the assumptions (a)-(c) hold.

If
\begin{equation}
\pi (t)\leq ce^{-at^{\gamma }},\ \ t\rightarrow \infty ;\ \ \ \pi (t)\leq
c/t^{\alpha /2},\ \ t\rightarrow 0  \label{c21}
\end{equation}
for some $\gamma >0$ and $\alpha \geq 0$, then for each $A>0$,
\begin{equation}
N_{0}(V)\leq C(h,A)[\int_{X_{h}^{-}}e^{-AW(x)^{-\gamma }}\mu
(dx)+\int_{X_{h}^{+}}bW(x)^{\max (\alpha /2,1)}\mu (dx)],  \label{max1}
\end{equation}
where $X_{h}^{-},~~X_{h}^{+},~~b$ are the same as in the theorem above.
\end{theorem}

\textbf{Proof of Theorem \ref{texp1}}. The proof is the same as that of the
theorem above. One only needs to replace (\ref{pow}) by the following
estimate
\begin{eqnarray*}
C(h)\int_{\frac{\sigma }{W}}^{\infty }e^{-at^{\gamma }}dt
&=&C(h)W^{-1}\int_{\sigma }^{\infty }e^{-a(\frac{\tau }{W})^{\gamma }}d\tau
\leq C(h)W^{-1}e^{-\frac{a}{2}(\frac{\sigma }{W})^{\gamma }}\int_{\sigma
}^{\infty }e^{-\frac{a}{2}(\frac{\tau }{W})^{\gamma }}d\tau  \\
&\leq &[C(h)W^{-1}\int_{\sigma }^{\infty }e^{-\frac{a}{2}(h\tau )^{\gamma
}}d\tau ]e^{-\frac{a}{2}(\frac{\sigma }{W})^{\gamma }},
\end{eqnarray*}
and note that $\sigma $ can be chosen as large as we please. \qed

\section{Low local dimension ($\alpha < 2.$)}

\textbf{1. Operators on lattices and groups.} It is easy to see
that Theorems \ref{texp1} and \ref{c1} are not exact if $\alpha
\leq 2$. We are going to illustrate this fact now and provide a
better result for the case $\alpha =0$ which occurs, for example,
when operators on lattices and discrete groups are considered. An
important example with $\alpha =1$ will be discussed in next
subsection (operators on quantum graphs).

Let $X=\{x\}$ be a countable set and $H_{0}$ be a difference operator on $%
L^{2}(X)$ which is defined by
\begin{equation}
(H_{0}\psi )(x)=\sum_{y\in X}a(x,y)\psi (y),  \label{22}
\end{equation}
where
\[
a(x,x)>0,~~a(x,y)=a(y,x)\leq 0,~~\sum_{y\in X}a(x,y)=0.
\]
A typical example of $H_{0}$ is the negative difference Laplacian on the
lattice $X=Z^{d}$, i.e.,
\begin{equation}
(H_{0}\psi)(x) =-\Delta \psi =\sum_{y\in Z^{d}:|y-x|=1}[\psi (x)-\psi (y],~~x\in
Z^{d},  \label{lo}
\end{equation}

We will assume that $0<a(x,x)\leq c_{0}<\infty $. Then Sp$H_{0}\subset \lbrack
0,2c_{0}]$. The operator $-H_{0}$ defines the Markov chain $x(s)$ on $X
$ with continuous time $s\geq 0$ which spends exponential time with
parameter $a(x,x)$ at each point $x\in X$ and then jumps to a point $y\in X$
with probability $r(x,y)=\frac{a(x,y)}{a(x,x)},~\sum_{y:y\neq x}r(x,y)=1$.
The transition matrix $p(t,x,y)=P_{x}(x_{t}=y)$ is the fundamental solution
of the parabolic problem
\[
\frac{\partial p}{\partial t}+H_{0}p=0,~~p(0,x,y)=\delta _{y}(x).
\]
Obviously, $p(t,x,x)\leq \pi (t)\leq 1$, and $\pi (t)\rightarrow 1$
uniformly in $x$ as $t\rightarrow 0$. The asymptotic behavior of $\pi (t)$
as $t\rightarrow \infty $ depends on operator $H_{0}$ and can be more or
less arbitrary.

Consider now the operator $H=H_{0}-m\delta _{y}(x)$ with the potential
supported on one point. The negative spectrum of $H$ contains at most one
eigenvalue (due to rank one perturbation arguments), and such an eigenvalue
exists if $m\geq c_{0}$. The latter follows from the variational principle,
since
\[
<H_{0}\delta _{y},~\delta _{y}>-m<\delta _{y},~\delta _{y}>~\leq ~c_{0}-m~<0.
\]
However, Theorems \ref{c1} and \ref{texp1} estimate the number of negative
eigenvalues $N(V)$ of the operator $H$ by $Cm$. Similarly, if
\[
V=-\sum_{1\leq i\leq n}m_{i}\delta (x-x_{i})
\]
and $m_{i}\geq c_{0}$, then $N(V)=n,$ but Theorems \ref{c1} and \ref{texp1}
give only that $N(V)\leq C\sum m_{i}.$ The following statement provides a
better result for the case under consideration than the theorems above. The
meaning of the statement below is that we replace $\max (\alpha /2,1)=1$ in (%
\ref{max}), (\ref{max1}) by $\alpha /2=0.$ Let us also mention that these
theorems can not be strengthened in a similar way if $0<\alpha \leq 2$ (see
Example \ref{ex}).

\begin{theorem}
\label{61} Let $H=H_{0}+V(x)$, where $H_{0}$ is defined in (\ref{22}), and
let assumptions of Theorem \ref{t1} hold. Then for each $h>0$,
\[
N_{0}(V)\leq C(h)[n(h)+\int_{0}^{\infty }\frac{\pi (t)}{t}\sum_{x\in
X_{h}^{-}}G(tW(x))dt],~~n(h)=\#\{x\in X_{h}^{+}\}.
\]
If, additionally, either (\ref{c2}) or (\ref{c21}) is valid for $\pi (t)$
as $t\rightarrow \infty ,$ then for each $A>0$,
\begin{equation}
N_{0}(V)\leq C(h)[\sum_{x\in X_{h}^{-}}W(x)^{\frac{\beta }{2}%
}+n(h)],~~n(h)=\#\{x\in X_{h}^{+}\} \label{1},
\end{equation}
\[
N_{0}(V)\leq C(h,A)[\sum_{x\in X_{h}^{-}}e^{-AW(x)^{-\gamma
}}+n(h)],~~n(h)=\#\{x\in X_{h}^{+}\},
\]
respectively.
\end{theorem}
\textbf{Remark}. Estimate (\ref{1}) for $N(V)$ in the case $X=Z^d$ can be found in \cite{1}.

\textbf{Proof}. In order to prove the first inequality, we split the
potential $V(x)=V_{1}(x)+V_{2}(x),$ where $V_{2}(x)=V(x)$ for $x\in
X_{h}^{+},$ $V_{2}(x)=0$ for $x\in X_{h}^{-}$. Now for each $\varepsilon \in
(0,1)$,
\begin{equation}
N_{0}(V)\leq N_{0}(\varepsilon ^{-1}V_{1})+N_{0}((1-\varepsilon
)^{-1}V_{2})=N_{0}(\varepsilon ^{-1}V_{1})+n(h).  \label{el}
\end{equation}
It remains to apply Theorem \ref{t1} to the operator $-\Delta +\varepsilon
^{-1}V_{1}$ and pass to the limit as $\varepsilon \rightarrow 1.$ The next
two inequalities follow from Theorems \ref{c1} and \ref{texp1}. \qed

\textbf{2. Operators on quantum graphs.} We will consider a
specific quantum graph $\Gamma ^{d},$ the so called
Avron-Exner-Last graph. Its vertices are the points of the lattice
$Z^{d},$ and the edges are all segments of length one connecting
neighboring vertices. Let $s\in \lbrack 0,1]$ be the natural
parameter on the edges (distance from one of the end points of the
edge). Consider the space $D$ of smooth functions $\varphi $ on
edges of $\Gamma ^{d}$ with the following (Kirchoff's) boundary
conditions at vertices: at each vertex $\varphi $ is continuous
and
\begin{equation}
\sum_{i=1}^{d}\varphi _{i}^{\prime }=0,  \label{sm}
\end{equation}
where $\varphi _{i}^{\prime }$ are the derivatives along the
adjoint edges in
the direction out of the vertex. The operator $H_{0}$ acts on functions $%
\varphi \in D$ as $-\frac{d^{2}}{ds^{2}}.$ The closure of this operator in $%
L^{2}(\Gamma ^{d})$ is a self-adjoint operator with the spectrum
$[0,\infty ) $ (see \cite{king})

\begin{theorem}
\label{tael} The assumptions of Theorems \ref{t1}, \ref{c1} hold for operator $%
H_0$ introduced in this section with the constants $\alpha, \beta
$ in Theorem \ref{c1} equal to $1$ and $d$, respectively.
\end{theorem}

One can easily see that there is a Markov process with the
generator $-H_{0}, $ and condition (a) of Theorem \ref{t1} holds.
In appendix 1, we'll estimate the function $p_{0}$ in order to
show that condition (b) holds and find constants $\alpha ,\beta $
defined in Theorem \ref{c1}. In fact, the same arguments can be
used to verify condition (a) analytically. \qed

As we discussed above, Theorem \ref{c1} is not exact if
$\alpha \leq 2 $. Theorem \ref{61} provides a better result in the
case $\alpha =0$. The situation is more complicated if $\alpha
=1.$ We will illustrate it using the operator $H_{0}$ on quantum
graph $\Gamma ^{d}$ defined above. We will consider two specific
classes of potentials. In one case, inequality (\ref {max}) is
valid with $\max (\alpha /2,1)=1$ replaced by $\alpha /2=1/2.$
However, inequality (\ref{max}) can not be improved for potentials
of the second type. The first class (regular potentials) consists
of piece-wise constant functions.

\begin{theorem}
Let $d\geq 3$ and $V$ be constant on each edge $e_{i}$ of the graph: $%
V(x)=-v_{i}<0,~x\in e_{i}.$ Then
\[
N_{0}(V)\leq c(h)(\sum_{i:~v_{i}\leq
h^{-1}}v_{i}^{d/2}+\sum_{i:~v_{i}>h^{-1}}\sqrt{v_{i}}).
\]
\end{theorem}

\textbf{Proof}. Put $V(x)=V_{1}(x)+V_{2}(x)$, where $V_{1}(x)=V(x)$ if $%
|V(x)|>h^{-1}$, $~V_{1}(x)=0$ if $|V(x)|\leq h^{-1}$. Then (see
(\ref{el}))
\[
N_{0}(V)\leq N_{0}(2V_{1})+N_{0}(2V_{2}).
\]
One can estimate $N(V_{1})$ from above (below) by imposing the
Neumann (Dirichlet) boundary conditions at all vertices of $\Gamma
.$ This leads to the estimates
\[
\sum_{i:~v_{i}>h^{-1}}\frac{\sqrt{2v_{i}}}{\pi }\leq N_{0}(V)\leq
\sum_{i:~v_{i}>h^{-1}}(\frac{\sqrt{2v_{i}}}{\pi }+1)\leq
c(h)\sum_{i:~v_{i}>h^{-1}}\sqrt{v_{i}},
\]
which, together with Theorem \ref{c1} applied to $N_{0}(2V_{2}),$
justifies the statement of the theorem. \qed

The same arguments allow one to get a more general result.

\begin{theorem}
Let $d\geq 3$. Let $\Gamma _{-}^{d}$ be the set of edges $e_{i}$
of the graph $\Gamma ^{d}$ where $W\leq h^{-1},~\Gamma _{+}^{d}$
be the complementary set of edges, and
\[
\frac{\sup_{x\in e_{i}}W(x)}{\min_{x\in e_{i}}W(x)}\leq
k_{0}=k_{0}(h),~x\in \Gamma _{+}^{d},
\]
where $W=V_{-}$. Then
\[
N_{0}(V)\leq c(h,k_{0})(\int_{\Gamma
_{-}^{d}}W(x)^{d/2}dx+\int_{\Gamma _{+}^{d}}\sqrt{W(x)}dx).
\]
\end{theorem}

%Note that the coefficient for the first integral above can be found in the
%proof of Corollary \ref{c1}, and the second term in the right hand side
%above estimates $m(\frac{2\sqrt{h}}{\pi }+1).$
\textbf{Example.}\label{ex} The next example shows that there are
singular potentials on $\Gamma ^{d}$ for which $\max (\alpha
/2,1)$ in (\ref{max})
can not be replaced by any value less than one. Consider the potential $%
V(x)=-A\sum_{i=1}^{m}\delta (x-x_{i})$, where $x_{i}$ are middle
points of some edges, and $A>4$. One can easily modify the example
by considering $\delta $ -sequences instead of $\delta $-functions
(in order to get a smooth potential.) Then
\[
\int_{\Gamma ^{d}}W^{\sigma }(x)dx=0
\]
for any $\sigma <1,$ while $N(V)\geq m.$ In fact, consider the
Sturm-Liouville problem on the interval $[-1/2,1/2]:$%
\[
-y^{\prime \prime }-A\delta (x)y=\lambda
y,~y(-1/2)=y(1/2)=0,~~A>4.
\]
It has (a unique) negative eigenvalue which is the root of the equation $%
\tanh (\sqrt{-\lambda }/2)=2\sqrt{-\lambda }/A.$ The corresponding
eigenfunction is $y=\sinh [\sqrt{-\lambda }(|x|+1/2)].$ The estimate $%
N(V)\geq m$ follows by imposing the Dirichlet boundary conditions
on the vertices of $\Gamma ^{d}.$

\section{\protect\bigskip Anderson model.}

\textbf{I. Discrete case.} Consider the classical Anderson Hamiltonian $%
H_{0}=-\Delta +V(x,\omega )$ on $L^{2}(Z^{d})$\ with random potential $%
V(x,\omega ).$ Here
\[
\Delta \psi (x)=\sum_{x^{\prime }:|x^{\prime }-x|=1}\psi (x^{\prime
})-2d\psi (x).
\]
We assume that random variables $V(x,\omega )$ on the probability space $(\Omega ,F,P)$
have the Bernoulli structure, i.e., they are i.i.d. and $P\{V(\cdot
)=0\}=p>0,$ $P\{V(\cdot )=1\}=q=1-p>0.$ The spectrum of $H_{0}$ is equal to
(see \cite{carl})
\[
\text{Sp}(H_{0})=\text{Sp}(-\Delta )\oplus 1=[0,4d+1].
\]

Let us stress that $0\in $Sp$(H_{0})$ due to the existence P-a.s. of
arbitrarily large clearings in realizations of $V$, i.e., there are balls
$B_{n}=\{x:|x-x_{n}|<r_{n}\}$ \ such that $V(x)=0,$ $x\in B_{n},$ and $%
r_{n}\rightarrow \infty $ as $n\rightarrow \infty $ (see the proof of the
theorem\ below for details).

Let
\[
H=H_{0}-W(x),~W(x)\geq 0.
\]
The operator $H$ has discrete random spectrum on $(-\infty ,0]$ with
possible accumulation point at $\lambda =0.$ Put $N_{0}(-W)=\#\{\lambda
_{i}\leq 0\}.$ Obviously, $N_{0}(-W)$ is random. Denote by $E$ the expectation of a r.v., i.e.
\[
EN_0=\int_{\Omega}N_0P(d\omega).
\]

\begin{theorem}
\label{ta}(a) For each $h>0\ $and $\gamma <\frac{d}{d+2},$%
\[
EN_{0}(-W)\leq c_{1}(h)[\#\{x\in Z^{d}:W(x)\geq h^{-1}\}]+c_{2}(h,\gamma
)\sum_{x:W(x)<h^{-1}}e^{-\frac{1}{W^{\gamma }(x)}}.
\]
In particular, if $W(x)<\frac{C}{\log ^{\sigma }|x|},$ $|x|\rightarrow
\infty ,$\ with some $\sigma >$ $\frac{d+2}{d},$ then $EN_{0}(-W)<\infty ,$
i.e., $N_{0}(-W)<\infty $ almost surely.

(b) If
\begin{equation}
W(x)>\frac{C}{\log ^{\sigma }|x|},\text{ }|x|\rightarrow \infty ,\ \text{\ \
and \ }\sigma <\frac{2}{d},  \label{log}
\end{equation}
then $N_{0}(-W)=\infty $ a.s. (in particular, $EN_{0}(-W)=\infty ).$
\end{theorem}

\textbf{Proof}. Since $V\geq 0,$ the kernel $p_{0}(t,x,y)$ of the semigroup $%
\exp(-tH_0)=\exp (t(\Delta -V))$ can be estimated by the kernel of $\exp (t\Delta ),$ i.
e., by the transition probability of the random walk with continuous time on
$Z^{d}$. The diagonal part of this kernel $p_{0}(t,x,x,\omega )$ is a
stationary field on $Z^{d}.$ Due to the Donsker-Varadhan estimate (see \cite
{DV},\cite{DV1}),
\[
Ep_{0}(t,x,x,\omega )=Ep_{0}(t,x,x,\omega )\stackrel{\log }{\sim }\exp
(-c_{d}t^{\frac{d}{d+2}}),\text{ \ \ }t\rightarrow \infty ,
\]
i.e.,
\[
\log Ep_{0}\sim -c_{d}t^{\frac{d}{d+2}},\text{ \ \ }t\rightarrow \infty .
\]
On the rigorous level, the relations above must be understood as
estimates from above and below, and the upper estimate has the following
form: for each $\delta >0,$
\begin{equation}
Ep_{0}\leq C(\delta )\exp (-c_{d}t^{\frac{d}{d+2}-\delta }),\text{ \ \ }%
t\rightarrow \infty .  \label{E}
\end{equation}
Now the first part of the theorem is a consequence of Theorems \ref{t1} and
\ref{texp1}. In fact, from Remarks \ref{r2} and \ref{r} and (\ref{E}) it
follows that
\begin{eqnarray*}
EN_{0}(V) &\leq &\frac{1}{c(\sigma )}\int_{X}W(x)\int_{\frac{\sigma }{W(x)}%
}^{\infty }Ep_{0}(t,x,x,\omega )dt\mu (dx) \\
&\leq &\frac{C(\delta )}{c(\sigma )}\int_{X}W(x)\int_{\frac{\sigma }{W(x)}%
}^{\infty }e^{-c_{d}t^{\frac{d}{d+2}-\delta }}dt\mu (dx).
\end{eqnarray*}
Then it only remains to repeat the arguments used to prove Theorem \ref
{texp1}.

The proof of the second part is based on the following lemma which indicates
the existence of large clearings at the distances which are not too
large. We denote by $C(r)$ the cube in the lattice,
\[
C(r)=\{x\in Z^{d}:|x_{i}|<r,\text{ }1\leq i\leq d\}.
\]
Let's divide $Z^{d}$ into cubic layers $L_{n}=C(a^{n+1})\backslash C(a^{n})$
\ with some constant $a\geq 1$ which will be selected later. One can choose
a set $\Gamma ^{(n)}=\{z_{i}^{(n)}\in L_{n}\}$ in each layer $L_{n}$ such
that
\[
|z_{i}^{(n)}-z_{j}^{(n)}|\geq 2n^{\frac{1}{d}}+1,\text{ \ \ \ }%
d(z_{i}^{(n)},\partial L_{n})>n^{\frac{1}{d}},
\]
and
\[
|\Gamma ^{(n)}|\geq c\frac{(2a)^{n(d-1)}a^{n+1}}{(2n^{1/d})^{d}}\geq ca^{nd},%
\text{ \ \ }n\rightarrow \infty .
\]
Let $C(n^{1/d},i)$ be the cube $C(n^{1/d})$ with the center shifted to the
point $z_{i}^{(n)}.$ Obviously, cubes $C_{n^{1/d},i}$ do not intersect each
other, $C(n^{1/d},i)\subset L_{n}$ and $|C(n^{1/d},i)|\leq c^{\prime }n.$

Consider the following event $A_{n}=$\{each cube $C(n^{1/d},i)\subset L_{n}$\
contains at least one point where $V(x)=1$\}. Obviously,
\[
P(A_{n})=(1-p^{|C(n^{1/d},i)|})^{|\Gamma ^{(n)}|}\leq e^{-|\Gamma
^{(n)}|p^{|C(n^{1/d},i)|}}\leq e^{-ca^{nd}c^{\prime
}p^{n}}=e^{-c(a^{d}p^{c^{\prime }})^{n}}.
\]

We will choose $a$ big enough, so that $a^{d}p^{c^{\prime }}>1.$ Then $\sum
P(A_{n})<\infty ,$ and the Borel-Cantelli lemma implies that $P$-a.s. there
exists $n_{0}(\omega )$ such that each layer $L_{n},$ $n\geq n_{0}(\omega ),$
contains at least one empty cube $C(n^{1/d},i),$ $i=i(n).$ $\ $Then from (%
\ref{log}) it follows that
\[
W(x)\geq \frac{C}{n^{\frac{2}{d}-\delta }}=\varepsilon _{n},\text{ \ \ \ }%
x\in C(n^{1/d},i),\text{ \ \ \ }i=i(n).
\]

One can easily show that the operator $H=-\Delta -\varepsilon $ in a cube $%
C\subset Z^{d}$ with the Dirichlet boundary condition at $\partial C$ has at
least one negative eigenvalue if $|C|\varepsilon ^{d/2}$ is big enough. Thus
the operator $H$ in $C(n^{1/d},i(n))$ with the Dirichlet boundary condition
has at least one eigenvalue if $n$ is big enough, and therefore $%
N(-W)=\infty .$ \qed

\textbf{II. Continuous case.} Theorem \ref{ta} is also valid for Anderson
operators in $R^{d}$. Let $H_{0}=-\Delta +V(x,\omega )$ on $L^{2}(R^{d})$\
with the random potential
\[
V(x,\omega )=\sum_{n\in Z^{d}}\varepsilon _{n}I_{Q_{n}}(x),~x\in
R^{d},~n=(n_{1},...,n_{d}),
\]
where $Q_{n}=\{x\in R^{d}:n_{i}\leq x_{i}<n_{i}+1,~i=1,2,...d\}$ and $%
\varepsilon _{n}$ are independent Bernoulli r.v. with $P\{\varepsilon
_{n}=0\}=p,~P\{\varepsilon _{n}=1\}=q=1-p.$ Put $H=H_{0}-W(x)=-\Delta
+V(x,\omega )-W(x).$

\begin{theorem}
\label{ta1}(a) If $d\geq 3$, then for each $h>0\ $and $\gamma <\frac{d}{d+2},
$
\[
EN_{0}(-W)\leq c_{1}(h)\int_{W(x)\geq h^{-1}}W(x)^{d/2}dx+c_{2}(h,\gamma
)\int_{W(x)<h^{-1}}e^{-\frac{1}{W^{\gamma }(x)}}dx.
\]
In particular, if $W(x)<\frac{C}{\log ^{\sigma }|x|},$ $|x|\rightarrow
\infty ,$\ with some $\sigma >$ $\frac{d+2}{d},$ then $EN_{0}(-W)<\infty ,$
i.e., $N_{0}(-W)<\infty $ almost surely.

(b) If
\[
W(x)>\frac{C}{\log ^{\sigma }|x|},\text{ }|x|\rightarrow \infty ,\ \text{\ \
and \ }\sigma <\frac{2}{d},
\]
then $N_{0}(-W)=\infty $ a.s. (in particular, $EN_{0}(-W)=\infty ).$
\end{theorem}

The proof of this theorem is identical to the proof of Theorem \ref{ta} with
the only difference that now $p_0(t,0,0)$ is not bounded as $t\rightarrow 0$%
, but $p_0(t,0,0)\leq c/t^{d/2},~t\rightarrow 0.$

\section{\protect\bigskip Lobachevsky plane,
processes with independent increments.}

\textbf{1. Lobachevsky plane }(see \cite{eis},
\cite{ras})\textbf{.} We will use the Poincare upper half plane
model, where $X=\{z=x+iy:y>0\}$ and the (Riemannian) metric on $X$
has the form
\begin{equation}
ds^{2}=y^{-2}(dx^{2}+dy^{2}).  \label{met}
\end{equation}
The geodesic lines of this metric are circular arcs perpendicular to the
real axis (half-circles whose origin is on the real axis) and straight
vertical lines ending on the real axis. The group of transformations
preserving $ds^{2}$ is $SL(2,R)$, i.e. the group of real valued $2\times 2$
matrices with the determinant equal to one. For each $A=\left[
\begin{array}{cc}
a & b \\
c & d
\end{array}
\right] \in SL(2,R),$ the action $A(z)$ is defined by
\[
A(z)=\frac{az+b}{cz+d}.
\]
For each $z_{0}\in X,$ there is a one-parameter stationary subgroup which consists of $A$ such that $Az_{0}=z_{0}$. The Laplace-Beltrami operator $\Delta ^{\prime }$
(invariant with respect to $SL(2,R)$) is defined uniquely up to a constant
factor, and is equal to
\begin{equation}
\Delta ^{\prime }=y^{2}\Delta =y^{2}(\frac{\partial ^{2}}{\partial x^{2}}+%
\frac{\partial ^{2}}{\partial y^{2}}),  \label{bl}
\end{equation}
The operator $-\Delta ^{\prime }$ is self-adjoint with respect to the
Riemannian measure
\begin{equation}
\mu (dz)=y^{-2}dxdy,  \label{mea}
\end{equation}
and has absolutely continuous spectrum on $[1/4,\infty ).$ In order to find
the number $N^{\prime }(V)$ of eigenvalues of the operator $-\Delta ^{\prime
}+V(x)$ below $1/4$, one can apply Theorem \ref{t1} to the operator $%
H_{0}=-\Delta ^{\prime }-\frac{1}{4}I.$

One needs to know constants $\alpha ,\beta $ in order to apply Theorem \ref
{c1}. It is shown in \cite{karp} that the fundamental solution for the
parabolic equation $u_{t}=-\Delta ^{\prime }u$ has the following asymptotic
behavior
\[
p(t,0,0)\sim c_{1}/t,~~t\rightarrow 0;~~p(t,0,0)\sim
c_{2}e^{-t/4}/t^{3/2},~~t\rightarrow \infty .
\]
Thus $\alpha =2,$ $\beta =3$ for the operator $H_{0}=-\Delta ^{\prime }-%
\frac{1}{4}I.$ A similar result for the Laplacian in the Hyperbolic space of
the dimension $d\geq 3$ can be found in \cite{1}.

\textbf{2. Markov processes with independent increments
(homogeneous pseudo differential operators).} We will estimate
$N_0(V)$ for shift invariant pseudo differential operators $H_0$
associated with Markov processes with independent increments.
Similar estimates were obtained in \cite{d} for pseudo
differential operators under assumptions that the symbol $f(p)$ of
the operator is monotone and non-negative, and the parabolic
semigroup $e^{-tH_0}$ is positivity preserving. This class
includes important cases of $f(p)=|p|^{\alpha}, \alpha <2$ and
$f(p)=\sqrt{p^2+m^2}-m$. Note that necessary and sufficient
conditions of the positivity of $p_0(t,x,x)$ are given by
Levy-Khinchin formula. We will omit monotonicity condition. What
is more important, the results will be expressed in terms of the
Levy measure responsible for the positivity of $p_0(t,x,x)$. This
will allow us to consider variety estimates with power and
logarithmical decaying potentials.

  Let $H_{0}$ be a pseudo-differential operator in $%
X=R^{d}$ of the form
\[
H_{0}u=F^{-1}\Phi (k)Fu,~~(Fu)(k)=\int_{R^{d}}u(x)e^{-i(x,k)}dx,~~~u\in
S(R^{d}),
\]
where the symbol $\Phi (k)$ of the operator $H_{0}$ has the following form
\begin{equation}
\Phi (k)=\int_{R^{d}}(1-\cos (x,k))\nu (x)dx.  \label{ef}
\end{equation}
Here $\mu (dx)=\nu (x)dx$ is an arbitrary measure (for simplicity we assumed
that it has a density) such that
\begin{equation}
\int_{|x|>1}\nu (x)dx+\int_{|x|<1}|x|^{2}\nu (x)dx<\infty .  \label{ef2}
\end{equation}
Assumption (\ref{ef}) is needed (and is sufficient) to construct a Markov
process with the generator $L=-H_{0}$ (see below). However, we will impose
an additional restriction on the measure $\mu (dx)$ assuming that the
density $\nu (x)$ has the following power asymptotics at zero and at
infinity
\[
\nu (x)\sim |x|^{-d-2+\rho },~~x\rightarrow 0,~~\nu (x)\sim |x|^{-d-\delta
},~~x\rightarrow \infty ,
\]
with some $\rho ,\delta \in (0,2).$ Note that assumption (\ref{ef2}) holds
in this case. To be more rigorous, we assume that
\begin{eqnarray}
\nu (x) &=&a(\frac{x}{|x|})|x|^{-d-\rho }(1+O(|x|^{\varepsilon
})),~~x\rightarrow 0,  \label{vu} \\
~~\nu (x) &=&b(\frac{x}{|x|})|x|^{-d-\delta }(1+O(|x|^{-\varepsilon
})),~~x\rightarrow \infty ,  \label{vu2}
\end{eqnarray}
where $a,b,\varepsilon >0.$ We also will consider another special case when
the asymptotic behavior of $\nu (x)$ at infinity is at logarithmical
borderline for the convergence of the integral (\ref{ef2}). Namely, we will
assume that (\ref{vu}) holds and
\begin{equation}
\nu (x)>C|x|^{-d}\log ^{-\sigma }|x|,~~x\rightarrow \infty ,\text{ \ \ }%
\sigma >1.  \label{vu3}
\end{equation}

The solution of problem (\ref{para}) is given by
\begin{equation}
p_{0}(t,x-y)=\frac{1}{(2\pi )^{d}}\int_{R^{d}}e^{-t\Phi (k)+i(x-y,k)}dk.
\label{p0}
\end{equation}

A special form of the pseudo differential operator $H_{0}$ is chosen in
order to guarantee that $p_{0}\geq 0.$ In fact, let $x_{s},$ $s>0,$ be a
Markov process in $R^{d}$ with symmetric independent increments. It means
that for arbitrary $0<s_{1}<s_{2}<...$ , the random variables $%
x_{s_{1}}-x_{0},$ $x_{s_{2}}-x_{s_{1}},...$ are independent and the
distribution of $x_{t+s}-x_{s}$ is independent of $s.$ The symmetry
condition means that Law$(x_{s}-x_{0})=$Law$(x_{0}-x_{s}),$ or $%
p(s,x,y)=p(s,y,x),$ where $p$ is the transition density of the process.
According to the Levy-Khinchin theorem (see \cite{gik}), the Fourier
transform (characteristic function) of this distribution has the form
\[
Ee^{i(k,x_{t+s}-x_{s})}=e^{-t\Phi (k)},
\]
with $\Phi (k)$ given by (\ref{ef}). Moreover, each measure (\ref{ef2})
corresponds to some process. One can consider the family of processes $%
x_{s}^{(x_{0})}=x_{0}+x_{s},$ $s>0,$ with an arbitrary initial point $%
x_{0}.\ $The generator $L$ of this family can be evaluated in the Fourier
space. If $\varphi (x)\in S(R^{d})$ and $\widehat{\varphi }(k)=F\varphi ,$
then
\begin{eqnarray*}
L\varphi (x) &=&\lim_{t\rightarrow 0}\frac{E\varphi (x+x_{t}^{(0)})-\varphi
(x)}{t}=\lim_{t\rightarrow 0}\frac{1}{(2\pi )^{d}}\int_{R^{d}}\frac{%
Ee^{i(x+x_{t}^{(0)},k)}-e^{i(x,k)}}{t}\widehat{\varphi }(k)dk \\
&=&\frac{-1}{(2\pi )^{d}}\int_{R^{d}}e^{i(x,k)}\Phi (k)\widehat{\varphi }%
(k)dk=-H_{0}\varphi .
\end{eqnarray*}
Thus, function (\ref{p0}) is the transition density of some process, and
therefore $p_{0}(t,x)\geq 0,$ i.e., assumption (a) of Theorem \ref{t1}
holds. Since operator $H_{0}$ is translation invariant, assumption (b)
also holds with $\pi (t)=p_{0}(t,0).$ Hence, Theorem \ref{t1} can be applied
to study negative eigenvalues of the operator $H_{0}+V(x)$ when (Levy)
measure $\nu dx$ satisfies (\ref{ef2})$.$ If (\ref{vu}), (\ref{vu2}) or (\ref
{vu}), (\ref{vu3}) hold, then Theorems \ref{c1}, \ref{texp1} can be used.
Namely, the following statement is valid.

\begin{theorem}
If measure $\nu dx$ satisfies (\ref{vu}) and (\ref{vu2}), then (\ref{c2}) is
valid with $\beta =2d/\delta ,$ $\alpha =2d/\rho .$

If measure $\nu dx$ satisfies (\ref{vu}) and (\ref{vu3}), then (\ref{c21})
is valid with $\gamma =1/\sigma ,$ $\alpha =2d/\rho .$
\end{theorem}

\textbf{Proof.} Consider first the case when (\ref{vu}) and (\ref{vu2})
hold. Let us prove that these relations imply the following behavior of $%
\Phi (k)$ at zero and at infinity
\begin{equation}
\Phi (k)=f(\frac{k}{|k|})|k|^{\delta }(1+O(|k|^{\varepsilon
_{1}})),~k\rightarrow 0;~\Phi (k)=g(\frac{k}{|k|})|k|^{\rho
}(1+O(|k|^{-\varepsilon _{1}})),~k\rightarrow \infty ,  \label{asy}
\end{equation}
with some $f,g,\varepsilon _{1}>0.$ We write (\ref{ef}) in the form
\begin{equation}
\Phi (k)=\int_{|x|<1}2\sin ^{2}(x,k))\nu (x)dx+\int_{|x|>1}2\sin
^{2}(x,k))\nu (x)dx=\Phi _{1}(k)+\Phi _{2}(k).  \label{asf}
\end{equation}
The term $\Phi _{1}(k)$ is analytic in $k$ and is of order $O(|k|^{2})$ as $%
k\rightarrow 0.$ We represent the second term as
\[
\int_{R^{d}}2\sin ^{2}(x,k))b(\stackrel{\cdot }{x})|x|^{-d-\delta
}dx-\int_{|x|<1}2\sin ^{2}(x,k))b(\stackrel{\cdot }{x})|x|^{-d-\delta
}dx+\int_{|x|>1}2\sin ^{2}(x,k))h(x)dx,
\]
where $\stackrel{\cdot }{x}=x/|x|$ and
\[
h(x)=\nu (x)-b(\stackrel{\cdot }{x})|x|^{-d-\delta },~~|h|\leq
C|x|^{-d-\delta -\varepsilon }.
\]
The middle term above is of order $O(|k|^{2})$ as $k\rightarrow 0.$ The
first term above can be evaluated by substitution $x\rightarrow x/|k|$. It
coincides with $f(\frac{k}{|k|})|k|^{\delta }.$ One can reduce $\varepsilon $
to guarantee that $\delta +\varepsilon <2$. Then the last term can be
estimated using the same \ substitution. This leads to the asymptotics (\ref
{asy}) as $k\rightarrow 0.$

Now let $|k|\rightarrow \infty .$ Since $\Phi _{2}(k)$ is bounded uniformly
in $k$, it remains to show that $\Phi _{1}(k)$ has the appropriate
asymptotics as $|k|\rightarrow \infty $. We write $v(x)$ in the integrand of
$\Phi _{1}(k)$ as follows
\[
v(x)=a(\stackrel{\cdot }{x})|x|^{-d-\rho }+g(x),~~|g(x)|\leq C|x|^{-d-\rho
+\varepsilon }.
\]
Then
\begin{eqnarray*}
\Phi _{1}(k) &=&\int_{R^{d}}2\sin ^{2}(x,k))a(\stackrel{\cdot }{x}%
)|x|^{-d-\rho }dx-\int_{|x|>1}2\sin ^{2}(x,k))a(\stackrel{\cdot }{x}%
)|x|^{-d-\rho }dx \\
&&+\int_{|x|<1}2\sin ^{2}(x,k))g(x)dx.
\end{eqnarray*}
The middle term in the right hand side above is bounded uniformly in $k$.
The substitution $x\rightarrow x/|k|$ justifies that the first term
coincides with $g(\frac{k}{|k|})|k|^{\rho }.$ The same substitution shows
that the order of the last term is smaller if $\varepsilon <\rho .$ This gives the second relation of (\ref
{asy}), and therefore, (\ref{asy}) is proved.

Let us estimate $\pi (t)$ when (\ref{asy}) holds$.$ From (\ref{p0}) it
follows that
\begin{equation}
\pi (t)=\frac{1}{(2\pi )^{d}}\int_{|k|<1}e^{-t\Phi (k)}dk+O(e^{-\eta t})~~%
\text{\textrm{as} \ }t\rightarrow \infty ,~~\eta >0.  \label{qqq}
\end{equation}
Now the substitution $k\rightarrow t^{-1/\delta }k$ leads to
\[
\pi (t)\sim ct^{-d/\delta },~~t\rightarrow \infty ,~~c=\frac{1}{(2\pi )^{d}}%
\int_{R^{d}}e^{-g(\frac{k}{|k|})|k|^{\delta }}dk.
\]
Hence, the first of relations (\ref{c2}) holds with $\beta =2d/\delta .$ In
order to estimate $\pi (t)$ as $t\rightarrow 0$, we put
\[
\pi (t)=\frac{1}{(2\pi )^{d}}\int_{|k|>1}e^{-t\Phi (k)}dk+O(1)~~\mathrm{as}%
~~t\rightarrow 0,
\]
and make the substitution $k\rightarrow t^{-1/\rho }k.$ This leads to
\[
\pi (t)\sim ct^{-d/\rho },~~t\rightarrow 0,~~c=\frac{1}{(2\pi )^{d}}%
\int_{R^{d}}e^{-f(\frac{k}{|k|})|k|^{\rho }}dk.
\]
Hence the second of relations (\ref{c2}) holds with $\alpha =2d/\rho .$ The
first statement of the theorem is proved.

Let us prove the second statement. If (\ref{vu}) and (\ref{vu3}) hold, then
\begin{equation}
\Phi (k)\geq c(\log \frac{1}{|k|})^{1-\sigma },~k\rightarrow 0;~\text{\ \ }%
\Phi (k)=g(\frac{k}{|k|})|k|^{\rho }(1+O(|k|^{-\varepsilon
_{1}})),~k\rightarrow \infty .  \label{asy1}
\end{equation}

In fact, only integrability of $v(x)$ at infinity, but not (\ref{vu2}), was
used in the proof of the second relation of (\ref{asy}). Thus the second
relation of (\ref{asy1}) is valid. Let us prove the first estimate. Let $%
\Omega _{k}=\{x:|k|^{-2}>|x|>|k|^{-1}\},$ $|k|<1.$ We have
\begin{eqnarray*}
\Phi (k) &\geq &\int_{\Omega _{k}}2\sin ^{2}(x,k))\nu (x)dx\geq
C\int_{\Omega _{k}}\sin ^{2}(x,k))|x|^{-d}\log ^{-\sigma }|x|dx \\
&\geq &C(2\log \frac{1}{|k|})^{-\sigma }\int_{\Omega _{k}}\sin
^{2}(x,k))|x|^{-d}dx,\text{ \ \ }|k|\rightarrow 0.
\end{eqnarray*}
It remains to show that
\begin{equation}
\int_{\Omega _{k}}\sin ^{2}(x,k))|x|^{-d}dx\sim \log \frac{1}{|k|},\text{ \
\ }|k|\rightarrow 0.  \label{qq}
\end{equation}
After the substitution $x=y/|k|,$ the last integral can be written in the
form
\[
\frac{1}{2}\int_{|k|^{-1}>|y|>1}|y|^{-d}dy-\frac{1}{2}\int_{|k|^{-1}>|y|>1}%
\cos (y,\stackrel{\cdot }{k}))|y|^{-d}dy.
\]
This justifies (\ref{qq}), since the second term above converges as $%
|k|\rightarrow 0.$ Hence (\ref{asy1}) is proved.

Finally, we need to obtain (\ref{c21})$.$ The estimation of $\pi (t)$ as $%
t\rightarrow 0$ remains the same as in the proof of the first statement of
the theorem. To get the estimate as $t\rightarrow \infty $, we use (\ref
{qqq}) (with a smaller domain of integration) and (\ref{asy1}). Then we
obtain
\[
\pi (t)\leq \frac{1}{(2\pi )^{d}}\int_{|k|<1/2}e^{-ct(\log \frac{1}{|k|}%
)^{1-\sigma }}dk+O(e^{-\eta t})~~\text{\textrm{as} \ }t\rightarrow \infty
,~~\eta >0.
\]
After integrating with respect to angle variables and substitution $\log
\frac{1}{|k|}=z$, we get
\[
\pi (t)\leq C\int_{\log 2}^{\infty }z^{d-1}e^{-z-ctz^{1-\sigma
}}dz+O(e^{-\eta t})~~\text{\textrm{as} \ }t\rightarrow \infty ,~~\eta >0.
\]
The asymptotic behavior of the last integral can be easily found using
standard Laplace method, and the integral behaves as $C_{1}t^{\frac{2d-1}{%
2d\sigma }}e^{-c_{1}t^{\frac{1}{\sigma }}}$ when $t\rightarrow \infty .$
This completes the proof of (\ref{c21}). \qed

\section{\protect\bigskip Continuous and discrete groups.}

\textbf{1. Free groups. }Let $X$ be a group $\Gamma $ with generators $%
a_{1}, $ $a_{2},$ ... $a_{d},$ inverse elements $a_{-1},$ $a_{-2},$ ... $%
a_{-d},$ the unit element $e$, and with no relations between generators
except $a_{i}a_{-i}=a_{-i}a_{i}=e$. The elements $g\in \Gamma $ are the shortest
versions of the words $g=$ $a_{i_{1}}\cdot ...\cdot a_{i_{n}}$ (with all
factors $e$ and $a_{j}a_{-j}$ being omitted). The metric on $\Gamma $ is
given by
\[
d(g_{1},g_{2})=d(e,g_{1}^{-1}g_{2})=m(g_{1}^{-1}g_{2}),
\]
where $m(g)$ is the number of letters $a_{\pm i}$ in $g.$ The measure $\mu $
on $\Gamma $ is defined by $\mu (\{g\})=1$ for each $%
g\in \Gamma .$ It is easy to see that $|\{g:d(e,g)=R\}|=2d(2d-1)^{R-1},$ i.e., the
group $\Gamma $ has an exponential growth rate.

Define the operator $\Delta _{\Gamma }$ on $X=\Gamma $ by the formula
\begin{equation}
\Delta _{\Gamma }\psi (g)=\mathrel{\mathop{\sum }\limits_{-d\leq i\leq
d,~i\neq 0}}[\psi (ga_{i})-\psi (g)].  \label{genlap}
\end{equation}
Obviously, the operator $-\Delta _{\Gamma }$ is bounded and non-negative in $%
L^{2}(\Gamma ,\mu )$. In fact, $||\Delta _{\Gamma }||\leq 4d.$ As it is easy
to see, the operator $\Delta _{\Gamma }$ is left-invariant:
\[
(\Delta _{\Gamma }\psi )(gx)=\Delta _{\Gamma }(\psi (gx)),~~x\in \Gamma ,
\]
for each fixed $g\in \Gamma .$ Thus, conditions (a), (b) hold for operator $%
-\Delta _{\Gamma }$. In order to apply Theorem \ref{c1}, one also needs to
find the parameters $\alpha $ and $\beta .$

\begin{theorem}
\label{tfree} a) The spectrum of the operator $-\Delta _{\Gamma }$ is
absolutely continuous and coincides with the interval $l_{d}=[\gamma ,\gamma
+4\sqrt{2d-1}],$ $\gamma =2d-2\sqrt{2d-1}\geq 0.$

b) The kernel of the parabolic semigroup $\pi _{\Gamma }(t)=(e^{t\Delta
_{\Gamma }})(t,e,e)$ on the diagonal has the following asymptotic behavior
at zero and infinity
\begin{equation}
\pi _{\Gamma }(t)\rightarrow c_{1}~~\mathrm{as}~~t\rightarrow 0,~\pi
_{\Gamma }(t)\sim c_{2}\frac{e^{-\gamma t}}{t^{3/2}}~~\mathrm{as}%
~~t\rightarrow \infty .  \label{asb}
\end{equation}
\end{theorem}

\begin{remark}
Since the absolutely continuous spectrum of the operator $-\Delta _{\Gamma }$
is shifted (it starts from $\gamma ,$ not from zero), the natural question
about the eigenvalues of the operator $-\Delta _{\Gamma }+V(g)$ is to
estimate the number $N_{\Gamma }(V)$ of eigenvalues below the threshold $%
\gamma .$ Obviously, $N_{\Gamma }(V)$ coincides with the number $N(V)$ of the
negative eigenvalues of the operator $H_{0}+V(g)$, where $H_{0}=-\Delta
_{\Gamma }-\gamma I.$ Hence one can apply Theorems \ref{t1}, \ref{61} to
this operator. From (\ref{asb}) it follows that constants $\alpha ,\beta $
for the operator $H_{0}=-\Delta _{\Gamma }-\gamma I$ are equal to $0$ and $3$%
, respectively, and
\[
N_{\Gamma }(V) \leq c(h)[n(h)+ \sum _{g \in \Gamma : W(g) \leq h^{-1}}
W(x)^{3/2}],~~n(h)=\# \{g \in \Gamma : W(g)>h^{-1} \}.
\]
\end{remark}

\textbf{Proof of Theorem \ref{tfree}}. Let us find the kernel $R_{\lambda
}(g_{1},g_{2})$ of the resolvent $(\Delta _{\Gamma }-\lambda )^{-1}.$ From
the $\Gamma $-invariance it follows that $R_{\lambda
}(g_{1},g_{2})=R_{\lambda }(e,g_{1}^{-1}g_{2}).$ Hence it is enough to
determine $u_{\lambda }(g)=R_{\lambda }(e,g)$. This function satisfies the
equation
\begin{equation}
\sum_{i\neq 0}u_{\lambda }(ga_{i})-(2d+\lambda )u_{\lambda }(g)=-\delta
_{e}(g),  \label{fre}
\end{equation}
where $\delta _{e}(g)=1$ if $g=e,$ $\delta _{e}(g)=0$ if $g\neq e.$ Since
the equation above is preserved under permutations of the generators, the
solution $u_{\lambda }(g)$ depends only on $m(g).$ Let $\psi _{\lambda }(m)=$
$u_{\lambda }(g),$ $m=m(g).$ Obviously, if $g\neq e$, then $m(ga_{i})=m(g)-1$
for one of the elements $a_{i},$ $i\neq 0,$ and $m(ga_{i})=m(g)+1$ for all
other elements $a_{i},$ $i\neq 0.$ Hence (\ref{fre}) implies
\begin{equation}
2d\psi _{\lambda }(1)-(2d+\lambda )\psi _{\lambda }(0)=-1,  \label{or}
\end{equation}
\[
\psi _{\lambda }(m-1)+(2d-1)\psi _{\lambda }(m+1)-(2d+\lambda )\psi
_{\lambda }(m)=0,~~~m>0.
\]

Two linearly independent solutions of these equations have the form $\psi
_{\lambda }(m)=\nu _{\pm }^{m},$ where $\nu _{\pm }$ are the roots of the
equation
\[
\nu ^{-1}+(2d-1)\nu -(2d+\lambda )=0.
\]
Thus
\[
\nu _{\pm }=\frac{2d+\lambda \pm \sqrt{(2d+\lambda )^{2}-4(2d-1)}}{2(2d-1)}.
\]
The interval $l_{d}$ was singled out as the set of real $\lambda $ such that
the discriminant above is not positive. Since $\nu _{+}\nu _{-}=1/(2d-1),$
we have
\[
|\nu _{\pm }|=\frac{1}{\sqrt{2d-1}}~\mathrm{for}~\lambda \in l_{d};~~~|\nu
_{+}|>\frac{1}{\sqrt{2d-1}},~~~|\nu _{-}|<\frac{1}{\sqrt{2d-1}}~~~\mathrm{for%
}~\mathrm{real}~\lambda \notin l_{d}.
\]
Now, if we take into account that the set $A_{m_{0}}=\{g\in \Gamma ,$ $%
m(g)=m_{0}\}$ has exactly $2d(2d-1)^{m_{0}-1}$ points, i.e., $\mu
(A_{m_{0}})=2d(2d-1)^{m_{0}-1},$ we get that
\begin{equation}
\nu _{-}^{m(g)}\in L^{2}(\Gamma ,\mu ),~\nu _{+}^{m(g)}\notin L^{2}(\Gamma
,\mu )~\mathrm{for}~\mathrm{real}~\lambda \notin l_{d},  \label{11}
\end{equation}
and
\begin{equation}
\int_{\Gamma \cap \{g:m(g)\leq m_{0}\}}|\nu _{\pm }|^{2m(g)}\mu (dg)\sim
m_{0}~\mathrm{as}~m_{0}\rightarrow \infty \mathrm{\ }~\mathrm{for}~\lambda
\notin l_{d}.  \label{12}
\end{equation}

Relations (\ref{11}) imply that $R\backslash l_{d}$ belongs to the resolvent
set of the operator $\Delta _{\Gamma }$ and that $R_{\lambda }(e,g)=c\nu
_{-}^{m(g)}.$ Relation (\ref{12}) implies that $l_{d}$ belongs to the
absolutely continuous spectrum of the operator $\Delta _{\Gamma }$ with
functions $(\nu _{+}^{m(g)}-\nu _{-}^{m(g)})$ being the eigenfunctions of
the continuous spectrum. Hence statement a) is justified.

Note that the constant $c$ in the formula for $R_{\lambda }(e,g)$ can be
found from (\ref{or}). This gives
\[
R_{\lambda }(e,g)=\frac{1}{(2d+\lambda )-2d\nu _{-}}\nu _{-}^{m(g)}.
\]
Thus
\[
R_{\lambda }(e,e)=\frac{1}{(2d+\lambda )-2d\nu _{-}}.
\]
Hence, for each $a>0,$%
\[
\pi _{\Gamma }(t)=\frac{1}{2\pi }\int_{a-i\infty }^{a+i\infty }e^{\lambda
t}R_{\lambda }(e,e)d\lambda =\frac{1}{2\pi }\int_{a-i\infty }^{a+i\infty
}e^{\lambda t}\frac{d\lambda }{(2d+\lambda )-2d\nu _{-}}.
\]
The integrand here is analytic with branching points at the ends of the
segment $l_{d},$ and the contour of integration can be bent into the left
half plane Re$\lambda <0$ and replaced by an arbitrary closed contour around $%
l_{d}.$ This immediately implies the first relation of (\ref{asb}). The
asymptotic behavior of the integral as $t\rightarrow \infty $ is defined by
the singularity of the integrand at the point $-\gamma $ (the right end of $%
l_{d}).$ Since the integrand there has the form $e^{\lambda t}[a+b\sqrt{%
\lambda +\gamma }+O(\lambda +\gamma )],$ $\lambda +\gamma \rightarrow 0,$\
this leads to the second relation of (\ref{asb}).

\textbf{2. General remark on left invariant diffusions on Lie
groups}. The examples below concern differential operators on the
continuous and discrete non-commutative groups $\Gamma $
(processes with independent increments considered in the previous
section are examples of operators on the
abelian groups $R^{d})$.

First we will consider the Heisenberg (nilpotent) group $\Gamma =H^{3}$ of
the upper triangular matrices
\begin{equation}  \label{heis}
g=\left[
\begin{array}{ccc}
1 & x & z \\
0 & 1 & y \\
0 & 0 & 1
\end{array}
\right],\left( x,y,z\right) \in R^{3},
\end{equation}
with units on the diagonal, and its discrete subgroup $ZH^{3}$, where $\left(
x,y,z\right) \in Z^{3}.$

Then we study (solvable) group of the affine transformations of the real
line: $x\rightarrow ax+b\,,a>0$, which has the matrix representation:
\[
Aff\left( R^{1}\right) =\left\{ g=\left[
\begin{array}{cc}
a & b \\
0 & 1
\end{array}
\right] ,\text{ }a>0,\text{ }(a,b)\in R^{2}\right\} ,
\]
and its subgroup generated by $\alpha _{1}=\left[
\begin{array}{cc}
e & e \\
0 & 1
\end{array}
\right] $ and $\alpha _{2}=\left[
\begin{array}{cc}
e & -e \\
0 & 1
\end{array}
\right] $ and their inverses $\alpha _{-1}=\left[
\begin{array}{cc}
e^{-1} & -1 \\
0 & 1
\end{array}
\right] $ and $\alpha _{-2}=\left[
\begin{array}{cc}
e^{-1} & 1 \\
0 & 1
\end{array}
\right] .$

There are two standard ways to construct the Laplacian on a Lie
group. A usual differential-geometric approach starts with the Lie algebra $%
\mathfrak{A}\Gamma$ on $\Gamma $, which can be considered either as the
algebra of the first order differential operators generated by the
differentiations along the appropriate one-parameter subgroups of $\Gamma
$, or simply as a tangent vector space $T\Gamma $ to $\Gamma $ at the unit
element $I$. The exponential mapping $\mathfrak{A}\Gamma \rightarrow \Gamma $
allows one to construct (at least locally) the general left invariant
Laplacian $\triangle _{\Gamma }$\ on $\Gamma $ as the image of the
differential operator $\sum_{ij}a_{ij}D_{i}D_{j}+\sum_{i}b_{i}D_{i}$ with
constant coefficients on $\mathfrak{A}\Gamma .$ The Riemannian metric $%
ds^{2}$ on $\Gamma $ and the volume element $dv$ can be defined now using
the inverse matrix of the coefficients of the Laplacian $\triangle _{\Gamma
}.$ It is important to note that additional symmetry conditions are needed
to determine $\triangle _{\Gamma }$ uniquely.

The central object in the probabilistic construction of the Laplacian (see,
for instance, McKean \cite{MK}) is the Brownian motion $g_{t\text{ }}$on $%
\Gamma $. We impose the symmetry condition $g_{t}\stackrel{law}{=}%
g_{t}^{-1}. $ Since $\mathfrak{A}\Gamma $ is a linear space, \ one can
define the usual Brownian motion $b_{t\text{ }}$on $\mathfrak{A}\Gamma $
with the generator $\sum_{ij}a_{ij}D_{i}D_{j}+\sum_{i}b_{i}D_{i}$. The
symmetry condition holds if $(I+db_{t})\stackrel{law}{=}(I+db_{t})^{-1}.$
The process $g_{t\text{ }}$(diffusion on $\Gamma )$ is given (formally) by
the stochastic multiplicative integral
\[
g_{t\text{ }}=\prod_{s=0}^{t}(I+db_{s}),
\]
or (more rigorously) by the Ito's stochastic differential equation
\begin{equation}
dg_{t\text{ }}=g_{t}db_{t}.  \label{dg}
\end{equation}
The Laplacian $\triangle _{\Gamma }$\ is defined now as the generator of the
diffusion:
\begin{equation}
\triangle _{\Gamma }f(g)=\lim_{\Delta t\rightarrow 0}\frac{Ef(g(I+b_{\Delta
t}))-f(g)}{\Delta t},\text{ \ }f\in C^{2}(\Gamma ).  \label{gen}
\end{equation}
The Riemannian metric form is defined as above (by the inverse matrix of the
coefficients of the Laplacian).

We will use the probabilistic approach to construct the Laplacian in the
examples below, since it allows us to easily incorporate the symmetry
condition.

\textbf{3. Heisenberg group }$\Gamma =H^{3}$ of the upper
triangular matrices (\ref{heis}) with units on the diagonal. We
have
\[
\mathfrak{A}\Gamma =\{A=\left[
\begin{array}{ccc}
0 & \alpha & \gamma \\
0 & 0 & \beta \\
0 & 0 & 0
\end{array}
\right] ,\text{ \ \ }(\alpha ,\beta ,\gamma )\in R^{3}\},\text{ \ \ \ \ }%
e^{A}=\left[
\begin{array}{ccc}
1 & \alpha & \gamma +\frac{\alpha \beta }{2} \\
0 & 1 & \beta \\
0 & 0 & 1
\end{array}
\right] .
\]
Thus $A\rightarrow \exp (A)$ is a one-to-one mapping of $\mathfrak{A}\Gamma $
onto $\Gamma .$ Consider the following Brownian motion on $\mathfrak{A}\Gamma
:$%
\[
b_{t}=\left[
\begin{array}{ccc}
0 & u_{t} & \sigma w_{t} \\
0 & 0 & v_{t} \\
0 & 0 & 0
\end{array}
\right] ,
\]
where $\sigma$ is a constant and $u_{t},$ $v_{t},$ $w_{t}$ are (standard) independent Wiener processes.
Then equation (\ref{dg}) has the form
\[
dg_{t}=\left[
\begin{array}{ccc}
0 & dx_{t} & dz_{t} \\
0 & 0 & dy_{t} \\
0 & 0 & 0
\end{array}
\right] =\left[
\begin{array}{ccc}
1 & x_{t} & z_{t} \\
0 & 1 & y_{t} \\
0 & 0 & 1
\end{array}
\right] \left[
\begin{array}{ccc}
0 & du_{t} & \sigma dw_{t} \\
0 & 0 & dv_{t} \\
0 & 0 & 0
\end{array}
\right] ,
\]
which implies that
\[
dx_{t}=du_{t},\text{ \ }dy_{t}=dv_{t},\text{ \ }dz_{t}=\sigma
dw_{t}+x_{t}dv_{t}.
\]
Under condition $g(0)=I,$ we get
\[
g_{t}=\left[
\begin{array}{ccc}
1 & u_{t} & \sigma w_{t}+\int_{0}^{t}u_{s}dv_{s} \\
0 & 1 & v_{t} \\
0 & 0 & 1
\end{array}
\right] .
\]
Let us note that the matrix
\[
(g_{t})^{-1}=\left[
\begin{array}{ccc}
1 & -u_{t} & u_{t}v_{t}-\sigma w_{t}-\int_{0}^{t}u_{s}dv_{s} \\
0 & 1 & -v_{t} \\
0 & 0 & 1
\end{array}
\right] =\left[
\begin{array}{ccc}
1 & -u_{t} & -\sigma w_{t}+\int_{0}^{t}v_{s}du_{s} \\
0 & 1 & -v_{t} \\
0 & 0 & 1
\end{array}
\right]
\]
has the same law as $g_{t.\text{ }}$Now from (\ref{gen}) it follows that
\[
(\Delta _{\Gamma }f)(x,y,z)=\frac{1}{2}[f_{xx}+f_{yy}+(\sigma
^{2}+x^{2})f_{zz}+2\sigma xf_{yz}].
\]
The matrix of the left invariant Riemannian metric has the form
\[
\left[
\begin{array}{ccc}
1 & 0 & 0 \\
0 & 1 & \sigma x \\
0 & \sigma x & \sigma ^{2}+x^{2}
\end{array}
\right] ^{-1}=\left[
\begin{array}{ccc}
1 & 0 & 0 \\
0 & \sigma ^{2}+x^{2} & -\sigma x \\
0 & -\sigma x & 1
\end{array}
\right] ,
\]
i.e.,
\[
ds^{2}=dx^{2}+(\sigma ^{2}+x^{2})dy^{2}+dz^{2}-2\sigma xdydz,\text{ \ \ \ \
\ }dV=dxdydz.
\]

Denote by $p_{\sigma }(t,x,y,z)$ the transition density for the process $%
g_{t}$\ (fundamental solution of the parabolic equation $u_{t}=\Delta
_{\Gamma }u).$ Let $\pi _{\sigma }(t)=p_{\sigma }(t,0,0,0).$

\begin{theorem}
\label{111} Function $\pi _{\sigma }(t)$ has the following asymptotic
behavior at zero and infinity:
\begin{equation}
\pi _{\sigma }(t)\sim \frac{c_0}{t^{3/2}},\text{ \ }t\rightarrow 0;\text{ \
\ }\pi _{\sigma }(t)\sim \frac{c}{t^{2}},\text{ \ }t\rightarrow
\infty,~~c=p_0(1,0,0),  \label{pp}
\end{equation}
i.e., Theorem \ref{c1} holds for operator $H=\Delta _{\Gamma }+V(x,y,z)$ with
$\alpha =3,$ $\beta =4.$
\end{theorem}

\textbf{Proof}. Since $H^{3}$ is a three dimensional manifold, the
asymptotics at zero is obvious. Let us prove the second relation of (\ref{pp}%
). We start with the simple case of $\sigma =0.$ The operator $\Delta
_{\Gamma }$ in this case is degenerate. However, the density $%
p_{0}(t,x,y,z) $ \ exists and can be found using H\"{o}rmander
hypoellipticity theory or by direct calculations. In fact, the joint
distribution of $(x_{t},y_{t},z_{t})$ is self-similar:
\[
(\frac{u_{t}}{\sqrt{t}},\frac{v_{t}}{\sqrt{t}},\frac{\int_{0}^{t}u_{s}dv_{s}%
}{t})=(u_{1},v_{1},\int_{0}^{1}u_{s}dv_{s}),
\]
i.e.,
\[
p_{0}(t,x,y,z)=\frac{1}{t^{2}}p_{0}(1,\frac{x}{\sqrt{t}},\frac{y}{\sqrt{t}},%
\frac{z}{t}),
\]
and therefore,
\[
p_{0}(t,0,0,0)=\frac{c}{t^{2}},\text{ \ \ }c=p_{0}(1,0,0,0).
\]

Let $\sigma ^{2}>0.$ Then
\[
p_{\sigma }(t,x,y,z)=\frac{1}{\sqrt{2\pi \sigma ^{2}t}}%
\int_{R^{1}}p_{0}(t,x,y,z_{1})e^{-\frac{(z-zz_{1})^{2}}{2\sigma ^{2}t}%
}dz_{1}.
\]
After rescaling $\frac{x}{\sqrt{t}}\rightarrow x,$ $\frac{y}{\sqrt{t}}%
\rightarrow y,\frac{z}{t}\rightarrow z,$ we get
\[
p_{\sigma }(t,x,y,z)=\frac{\sqrt{t}}{t^{2}\sqrt{2\pi \sigma ^{2}}}%
\int_{R^{1}}p_{0}(1,x,y,z_{1})e^{-\frac{t(z-zz_{1})^{2}}{2\sigma ^{2}}%
}dz_{1}.
\]
From here it follows that $p_{\sigma }(t,0,0,0)\sim c/t^{2},$ $t\rightarrow
\infty ,$ with \ $c=p_{0}(1,0,0,0).$ \qed

Theorem \ref{111} can be proved for the group $H^n$ of $n\times n$ upper
triangular matrices with units on the diagonal. In this case,
\[
\alpha=\text{dim}H^n=\frac{n(n-1)}{2},~~\beta=(n-1)+2(n-2)+3(n-3)+...=\frac{%
n(n^2-1)}{2}.
\]

\bigskip \textbf{4. Heisenberg discrete group }$\Gamma =ZH^{3}$ of
integer valued matrices of the form
\[
g=\left(
\begin{array}{ccc}
1 & x & y \\
0 & 1 & z \\
0 & 0 & 1
\end{array}
\right) ,~~x,y,z\in Z^{1}.
\]

Consider the Markov process $g_{t}$ on $ZH^{3}$ defined by the equation
\begin{equation}
g_{t+dt}=g_{t}\left(
\begin{array}{ccc}
1 & d\xi _{t} & d\zeta _{t} \\
0 & 1 & d\eta _{t} \\
0 & 0 & 1
\end{array}
\right) ,  \label{eq12}
\end{equation}
where $\xi _{t},\eta _{t},\zeta _{t}$ are three independent Markov processes
on $Z^{1}$ with generators
\[
\Delta _{1}\psi (n)=\psi (n+1)+\psi (n-1)-2\psi (n),\text{ \ \ \ }n\in
Z^{1}.
\]
Equation (\ref{eq12}) can be solved using discretization of time. This
gives
\[
g_{t}=\left(
\begin{array}{ccc}
1 & x_{t} & y_{t} \\
0 & 1 & z_{t} \\
0 & 0 & 1
\end{array}
\right) \left(
\begin{array}{ccc}
1 & \xi _{t} & \zeta _{t}+\int_{0}^{t}\xi _{s}d\eta _{s} \\
0 & 1 & \eta _{t} \\
0 & 0 & 1
\end{array}
\right) .
\]

The generator $L$ of this process has the form (\ref{genlap}) with
\[
a_{\pm 1}=\left(
\begin{array}{ccc}
1 & \pm 1 & 0 \\
0 & 1 & 0 \\
0 & 0 & 1
\end{array}
\right) ,\text{ }a_{\pm 2}=\left(
\begin{array}{ccc}
1 & 0 & 0 \\
0 & 1 & \pm 1 \\
0 & 0 & 1
\end{array}
\right) ,\text{ }a_{\pm 3}=\left(
\begin{array}{ccc}
1 & 0 & \pm 1 \\
0 & 1 & 0 \\
0 & 0 & 1
\end{array}
\right),
\]
i.e.,
\begin{equation}
L=\Delta _{\Gamma }\psi (g)=\sum _{i=\pm1,\pm2,\pm3}[\psi (ga_{i})-\psi (g)].
\label{genlap1}
\end{equation}
If $\psi=\psi(g)$ is considered as a function of $(x,y,z)\in Z^3$, then
\begin{eqnarray}
L\psi (x,y,z) &=&\psi (x+1,y,z)+\psi (x-1,y,z)+\psi (x,y+1,z+x)+\psi
(x,y-1,z-x)  \nonumber \\
&& + \psi (x,y,z+1)+\psi (x,y,z-1)-6\psi (x,y,z).  \label{dgaz}
\end{eqnarray}

The analysis of the transition probability in this case is similar to the
continuous case, and it leads to the following result

\begin{theorem}
If $g_{t}$ is the process on $ZH^{3}$ with the generator (\ref{dgaz}), then
\[
P\{g_{t}=I\}=P\{x_{t}=y_{t}=z_{t}=0\}\sim \frac{c}{t^{2}},\text{ }%
t\rightarrow \infty,
\]
with $c$ defined in (\ref{pp}). In particular, Theorem \ref{61}\ can be
applied to operator $H_{0}=L$ with $\beta =4.$
\end{theorem}

This result is valid in a more general setting (see \cite{kmm}). Consider
three independent processes $\xi _{t},\eta _{t},\zeta _{t}$, $t\geq 0,$ on $%
Z^{1}$ with independent increments and such that
\begin{eqnarray*}
Ee^{ik\xi _{t}} &=&e^{-t(1-\sum_{i=1}^{\infty }p_{i}\cos ki)},\text{ }%
\sum_{i=1}^{\infty }p_{i}=1, \\
Ee^{ik\eta _{t}} &=&e^{-t(1-\sum_{i=1}^{\infty }q_{i}\cos ki)},\text{ }%
\sum_{i=1}^{\infty }q_{i}=1, \\
Ee^{ik\zeta _{t}} &=&e^{-t(1-\sum_{i=1}^{\infty }r_{i}\cos ki)},\text{ }%
\sum_{i=1}^{\infty }r_{i}=1.
\end{eqnarray*}
Assume also that there exist $\alpha _{1},\alpha _{2},\alpha _{3}$ on the
interval $(0,2)$ such that
\[
p_{i}\sim \frac{c_{1}}{i^{1+\alpha _{1}}},\text{ \ }q_{i}\sim \frac{c_{2}}{%
i^{1+\alpha _{2}}},\text{ \ }r_{i}\sim \frac{c_{3}}{i^{1+\alpha _{3}}}
\]
as $i\rightarrow \infty ,$ i.e., distributions with characteristic functions $%
\sum_{i=1}^{\infty }p_{i}\cos ki,$ $\sum_{i=1}^{\infty }q_{i}\cos ki,$ $%
\sum_{i=1}^{\infty }r_{i}\cos ki$ belong to the domain of attraction of the
symmetric stable law with parameters $\alpha _{1},\alpha _{2},\alpha _{3}.$
Let $g_{t}$ be the process on $ZH^{3}$ defined by (\ref{eq12}). Then
\[
P\{g_{t}=I\}\sim \frac{c}{t^{\gamma }},\text{ }t\rightarrow \infty \text{,
\ }\gamma =\max (\frac{2}{\alpha _{1}}+\frac{2}{\alpha _{1}},~\frac{1}{%
\alpha _{3}}).
\]

\textbf{5. Group $Aff\left( R^{1}\right)$ of affine
transformations of the real line.} This group of transformations
$x\rightarrow ax+b\,,$ $a>0,$ has a matrix representation:
\[
\Gamma =Aff\left( R^{1}\right) =\{g=\left[
\begin{array}{cc}
a & b \\
0 & 1
\end{array}
\right] ,\text{ \ }a>0,\text{ \ }(a,b)\in R^{2}\}.
\]

We start with the Lie algebra for $Aff\left( R^{1}\right) :$%
\[
\mathfrak{A}\Gamma =\left\{ \left[
\begin{array}{cc}
\alpha & \beta \\
0 & 0
\end{array}
\right] ,\text{ \ }\left( \alpha ,\beta \right) \in R^{2}\right\} .
\]

Obviously, for arbitrary $A=\left[
\begin{array}{cc}
\alpha & \beta \\
0 & 0
\end{array}
\right] ,$ one has
\[
\exp \left( A\right) =\left[
\begin{array}{cc}
e^{\alpha } & \beta \frac{e^{\alpha }-1}{\alpha } \\
0 & 1
\end{array}
\right] ,
\]
i.e., the exponential mapping of $\mathfrak{A}\Gamma $ coincides with the
group $\Gamma .$ Consider the diffusion
\[
b_{t}=\left[
\begin{array}{cc}
w_{t}+\alpha t & v_{t} \\
0 & 0
\end{array}
\right]
\]
on $\mathfrak{A}\Gamma $, where $\left( w_{t},v_{t}\right) $ are independent
Wiener processes. Consider the matrix valued process $g_{t}=\left[
\begin{array}{cc}
x_{t} & y_{t} \\
0 & 1
\end{array}
\right] ,\ g_{0}=\left[
\begin{array}{cc}
1 & 0 \\
0 & 1
\end{array}
\right] ,\ $on $\Gamma $ satisfying the equation
\[
dg_{t}=g_{t}db_{t}=\left[
\begin{array}{cc}
x_{t} & y_{t} \\
0 & 1
\end{array}
\right] \left[
\begin{array}{cc}
dw_{t}+\alpha dt & dv_{t} \\
0 & 0
\end{array}
\right] =\left[
\begin{array}{cc}
x_{t}\left( dw_{t}+\alpha dt\right) & x_{t}dv_{t} \\
0 & 0
\end{array}
\right] .
\]
This implies
\begin{eqnarray*}
dx_{t} &=&x_{t}\left( dw_{t}+\alpha dt\right) , \\
dy_{t} &=&x_{t}dv_{t},
\end{eqnarray*}
i.e. (due to Ito's formula),
\[
x_{t}=e^{w_{t}+\left( \alpha -\frac{1}{2}\right) t},\text{ \ }%
y_{t}=\int_{0}^{t}x_{s}dv_{s}.
\]

We impose the following symmetry condition:
\begin{equation}
\left( g_{t}\right) ^{-1}\stackrel{law}{=}g_{t},  \label{symm}
\end{equation}
It holds if $\alpha =\frac{1}{2}.$ In fact,
\begin{equation}  \label{1213}
g_{t}=\left[
\begin{array}{cc}
e^{w_{t}} & \int_{0}^{t}e^{w_{s}}dv_{s} \\
0 & 1
\end{array}
\right] ,\text{ \ }g_{t}^{-1}=\left[
\begin{array}{cc}
e^{-w_{t}} & -\int_{0}^{t}e^{w_{s}-w_{t}}dv_{s} \\
0 & 1
\end{array}
\right],
\end{equation}
and (\ref{symm}) follows after the change of variables $s=t-\tau $ in the
matrix $g_{t}^{-1}.$ Then the generator of the process $g_{t}$ has the form

\[
\triangle _{\Gamma}f=\frac{x^{2}}{2}\left[ \frac{\partial ^{2}f}{\partial
x^{2}}+\frac{\partial ^{2}f}{\partial y^{2}}\right] +\frac{x}{2}\frac{%
\partial f}{\partial x}.
\]

\begin{theorem}
Operator $\triangle _{\Gamma }$ is self-adjoint with respect to the measure $%
x^{-1}dxdy.$ The function $\pi \left( t\right) =p\left( t,0,0\right) $ $\ $%
has the following behavior at zero and infinity$:$%
\begin{equation}
\pi \left( t\right) \sim \frac{c_{0}}{t},\text{ }t\rightarrow 0;\text{ \ \ }%
\pi \left( t\right) \sim \frac{C}{t^{3/2}},\text{ \ }t\rightarrow \infty
\text{\ .}  \label{45}
\end{equation}
\end{theorem}
\begin{remark}Let $H=\triangle _{\Gamma}+V,$ where the negative part $W=V_{-}$ of the
potential is bounded: $W\leq h^{-1}.$ From (\ref{45}) and Theorem \ref{c1}
it follows that
\[
N_{0}(V)\leq C(h)\int_{-\infty }^{\infty }\int_{0}^{\infty }\frac{%
W^{3/2}(x,y)}{x}dxdy.
\]
\end{remark}
\begin{remark}The left-invariant Riemannian metric on $Aff\left(
R^{1}\right) $ is given by the inverse diffusion matrix of $%
\triangle_{\Gamma}$, i.e.,
\[
d\xi ^{2}=x^{-2}\left( dx^{2}+dy^{2}\right) \text{ \ \ \ \ }\left( g=\left[
\begin{array}{cc}
x & y \\
0 & 1
\end{array}
\right] ,\text{ \ \ }x>0\right)
\]
After the change $(x,y)\rightarrow (y,x),$ this formula coincides with the
metric on the Lobachevsky plane (see the previous section). However, one can not
identity the Laplacian on $Aff\left( R^{1}\right) $ and on the Lobachevsky
plane $L^{2}$, since they are defined by different symmetry conditions. The
plane $L^{2}$ has a three dimensional group of transformations, and each
point $z\in L^{2}$ has a one-parameter stationary subgroup. The Laplacian on the Lobachevsky plane was defined by the
invariance with respect to this three dimensional group of transformations.
In the case of $\Gamma =Aff\left( R^{1}\right) ,$ the group of
transformations is two dimensional. It acts as a left shift $g\rightarrow
g_{1}g,$ $g_{1},g\in \Gamma ,$ and the Laplacian is specified by the left
invariance with respect to this two dimensional group and the symmetry
condition (\ref{symm}).
\end{remark}
\textbf{Proof. }Since $\Gamma$ is a two dimensional manifold, the asymptotics
of $\pi (t)$ at zero is obvious. One needs only to justify the asymptotics
of $\pi \left( t\right) $ at infinity.

Let's find the density of $\left( x_{t},y_{t}\right)
=(e^{w_{t}},\int_{0}^{t}e^{w_{s}}dv_{s})$. The second term, for a fixed realization of $%
w_{\cdot }$, has the Gaussian law with (conditional) variance $%
\sigma ^{2}=\int_{0}^{t}e^{2w_{s}}ds,$ and
\begin{equation} \label{star}
P\ \{x_{t}\in 1+dx,\text{ }y_{t}\in 0+dy\}=p(t,0,0)dxdy=\frac{1}{\sqrt{2\pi t%
}}E\frac{1}{\sqrt{2\pi \int_{0}^{t}e^{2\hat{w}_{s}}ds}}.
\end{equation}
Here $\hat{w}_{s},$ $s\in \left[ 0,t\right] ,$ is the Brownian bridge on $%
\left[ 0,t\right] .$ The distribution of the exponential functional $A\left(
t\right) =\int_{0}^{t}e^{2\hat{w}_{s}}ds$ and the joint distribution of $%
\left( A\left( t\right) ,w\left( t\right) \right) $ were calculated in \cite
{Yor}. Together with (\ref{star}), these easily imply the statement of the theorem.
\qed

\textbf{6. A relation between Markov processes and random walks on
discrete groups}. Let $\Gamma $ be a discrete group generated by elements $%
a_{1},...,a_{d},$ $a_{-1}=a_{1}^{-1},...,a_{-d}=a_{d}^{-1},$ with some
identities. Define the Laplacian on $\Gamma $ by the formula
\[
\Delta \psi (g)=\sum_{i=-d}^{d}\psi (ga_{i})-2d\psi (g),\text{ \ \ }g\in
\Gamma .
\]
Consider the Markov process $g_{t}$ on $\Gamma $ with continuous time and
the generator $\Delta .$ Let $\widetilde{g}_{k},$ $k=0,1,2,...,$\ be the
Markov chain on $\Gamma $ with discrete time (symmetric random walk) such
that
\[
P\{\widetilde{g}_{0}=e\}=1,\text{ \ \ }P\{\widetilde{g}_{n+1}=ga_{i}\text{ }|%
\text{\ }\widetilde{g}_{n}=g\}=\frac{1}{2d},\text{ \ }i=\pm 1,\pm 2,...\pm
d.
\]
Then there is a relation between transition probability $p(t,e,g)$ of the
Markov process $g_{t}$ and the transition probability $P\{\widetilde{g}%
_{k}=g $ $\}$ of the random walk. In particular, one can estimate $\pi (t)=$
$p(t,e,e)$ for large $t$ through $\widetilde{\pi }(2k)=P\{\widetilde{g}%
_{2k}=e\}$ under minimal assumptions on $\widetilde{\pi }(2k).$ For example,
it is enough to assume that $\widetilde{\pi }(2k)=k^{\gamma}L(k),~\gamma
\geq 0$, where $L(k)$ for large $k$ can be extended as slowly varying monotonic
function of continuous argument $k.$ We are not going to provide a general
statement of this type, but we restrict ourself to a specific situation
needed in the next section. Note that we consider here only even arguments
of $\widetilde{\pi },$ since $\widetilde{\pi }(2k+1)=0.$

\begin{theorem}
\label{tll}Let
\[
\widetilde{\pi }(2n)\leq e^{-c_{0}(2n)^{\alpha }},~ n\rightarrow \infty ,~
c_{0}>0,~0<\alpha <1.
\]
Then
\[
\pi (t)\leq e^{-c_{0}(2dt)^{\alpha }},~t\geq t_{0}.
\]
\end{theorem}

\textbf{Proof. }The number $\nu _{t}$ of jumps of the process $g_{t}$ on the
interval $(0,t)$ has Poisson distribution. At the moments of jumps, the
process performs the symmetric random walk with discrete time and transition
probabilities $P\{g\rightarrow ga_{i}\}=1/2d,$ $i=\pm 1,\pm 2,...\pm d.$
Thus (taking into account that $\widetilde{\pi }(2k+1)=0$),
\[
\pi (t)=p(t,e,e)=\sum_{n=0}^{\infty }\widetilde{\pi }(2n)P\{\nu _{t}=2n\}.
\]

Due to the exponential Chebyshev inequality
\[
P\{|\nu _{t}-2dt|\geq \varepsilon t\}\leq e^{-c\varepsilon ^{2}t},\text{ \ }%
t\rightarrow \infty .
\]
Secondly,
\[
P\{\nu _{t}\text{ is even}\}=\frac{1}{2}+O(e^{-4dt}),\text{ \ }t\rightarrow
\infty .
\]
These relations imply that, for $t\rightarrow \infty $ and $\delta >0,$%
\begin{eqnarray*}
\pi (t) &=&\sum_{n:|2n-2dt|<\varepsilon t}\widetilde{\pi }(2n)P\{\nu
_{t}=2n\}+O(e^{-c_{0}(2dt)^{\alpha }}) \\
&\leq &\sum_{n:|2n-2dt|<\varepsilon t}e^{-c_{0}(2n)^{\alpha }}P\{\nu
_{t}=2n\}+O(e^{-c_{0}(2dt)^{\alpha }}) \\
&\leq &(1+\delta )e^{-c_{0}(2dt)^{\alpha }}\sum_{n:|2n-2dt|<\varepsilon
t}P\{\nu _{t}=2n\}+O(e^{-c_{0}(2dt)^{\alpha }}) \\
&\leq &\frac{1+\delta }{2}e^{-c_{0}(2dt)^{\alpha }}+O(e^{-c_{0}(2dt)^{\alpha
}}).
\end{eqnarray*}

\textbf{7. Random walk on the discrete subgroup of }$Aff\left( R^{1}\right) $%
\textbf{.} Let us consider the following two matrices $\alpha _{1}=\left[
\begin{array}{cc}
e & e \\
0 & 1
\end{array}
\right] $ and $\alpha _{2}=\left[
\begin{array}{cc}
e & -e \\
0 & 1
\end{array}
\right] $ in $Aff\left( R^{1}\right) $ and their inverses $\alpha _{-1}=\left[
\begin{array}{cc}
e^{-1} & -1 \\
0 & 1
\end{array}
\right] $ and $\alpha _{-2}=\left[
\begin{array}{cc}
e^{-1} & 1 \\
0 & 1
\end{array}
\right] .$ Let $G$ be a subgroup of $Aff\left( R^{1}\right) $ generated by $%
\alpha _{\pm 1}$and $\alpha _{\pm 2}.$ Consider the random walk on $G$ of
the form
\[
g_{n}=h_{1}h_{2}...h_{n},
\]
where one step random matrices $h_{i}$ coincide with one of the matrices $%
\alpha _{\pm 1},$ $\alpha _{\pm 2}$ with probability $1/4,$ i.e.,
\[
h_{i}=\left[
\begin{array}{cc}
e^{\varepsilon _{i}} & \delta _{i} \\
0 & 1
\end{array}
\right] ,
\]
where
\begin{eqnarray}
P\{\varepsilon _{i} &=&1,\text{ }\delta _{i}=e\}=P\{\varepsilon _{i}=1,\text{
}\delta _{i}=-e\}  \nonumber \\
&=&P\{\varepsilon _{i}=-1,\text{ }\delta _{i}=-1\}=P\{\varepsilon _{i}=-1,%
\text{ }\delta _{i}=1\}=1/4.  \label{err}
\end{eqnarray}

Let $\Delta_{G}$ be the Laplacian on $G$ which corresponds to the generators
$a_{\pm1},a_{\pm2}$, i.e., (compare with (\ref{genlap})(\ref{genlap1}))
\[
L=\Delta _{\Gamma }\psi (g)=\sum _{i=\pm1,\pm2}[\psi (ga_{i})-\psi (g)].
\]

\begin{theorem}
(a) The following estimate is valid for $\widetilde{\pi }(2n):$
\[
\widetilde{\pi }(2n)\leq e^{-c_{0}(2n)^{1/3}},~n\rightarrow \infty ,c_{0}>0.
\]

(b) Theorem \ref{61} can be applied to operator $H=\Delta _{G}+V(g)$ with $%
\gamma =1/3$, i.e.,
\[
N_{0}(V)\leq C(h,A)[\sum_{g:V(g)\leq
h^{-1}}e^{-AW(g)^{-1/3}}+n(h)],~~n(h)=\#\{g:W(g)>h^{-1}\}.
\]
\end{theorem}

\textbf{Proof. } The random variables $(\varepsilon _{i},$ $\delta _{i})$ are dependent,
but (\ref{err}) implies that $(\varepsilon _{i},$ $\widetilde{\delta }_{i}),$
where $\widetilde{\delta }_{i}=$sgn$\delta _{i},$ are independent symmetric
Bernoulli r.v. It is easy to see that
\[
g_{n}=\left[
\begin{array}{cc}
e^{S_{n}} & \sum_{k=1}^{n}\delta _{k}e^{S_{k-1}} \\
0 & 1
\end{array}
\right] ,
\]
where $S_{0}=1,$ $S_{k}=\varepsilon _{1}+...+\varepsilon _{k},$ $k>0,$ is
a symmetric random walk on $Z^{1}.$ This formula is an obvious discrete
analogue of (\ref{1213}). Our goal is to calculate the probability
\begin{eqnarray*}
\widetilde{\pi }(2n) &=&P\{g_{2n}=I\}=P\{S_{2n}=0,\text{ }%
\sum_{k=1}^{2n}\delta _{k}e^{S_{k-1}}=0\}=\left(
\begin{array}{c}
2n \\
n
\end{array}
\right) \frac{1}{2^{2n}}P\{\sum_{k=2}^{2n}\delta _{k}e^{\widehat{S}%
_{k-1}}=0\} \\
&\sim &\frac{1}{\sqrt{\pi n}}P\{\sum_{k=1}^{2n-1}\delta _{k+1}e^{\widehat{S}%
_{k}}=0\},\text{ \ }n\rightarrow \infty .
\end{eqnarray*}
Here $\widehat{S}_{k}$, $k=0,1,...,2n,$ is the discrete bridge, i.e., the
random walk $S_{k}$ under conditions $S_{0}=S_{2n}=0.$

Put $M_{2n}=\mathrel{\mathop{\max }\limits_{k\leq 2n}}\widehat{S}_{k},$ $%
m_{2n}=\mathrel{\mathop{\min }\limits_{k\leq 2n}}\widehat{S}_{k}.$ Let $%
\Gamma _{s-1}^{+},$ $\Gamma _{s}^{-}$ be the sets of moments of time $k$\
when the bridge $\widehat{S}_{k}$ changes value from $s-1$ to $s$ or
from $s$ to $s-1,$ respectively. Introduce local times $\tau _{s-1}^{+}=$%
Card$\Gamma _{s-1}^{+}$ and $\tau _{s}^{-}=$Card$\Gamma _{s}^{-}$, i.e. $%
\tau _{s-1}^{+}=\#($jumps of $\widehat{S}_{k}$ from $s-1$ to $s)$ and $\tau
_{s}^{-}=\#($jumps of $\widehat{S}_{k}$ from $s$ to $s-1).$ Note that $%
\delta _{k+1}e^{\widehat{S}_{k}}=\widetilde{\delta }_{k+1}e^{s}$ when $k\in
\Gamma _{s-1}^{+}\cup \Gamma _{s}^{-},$ and therefore
\[
\sum_{k=1}^{2n-1}\delta _{k+1}e^{\widehat{S}_{k}}=%
\sum_{s=m_{2n}+1}^{M_{2n}}e^{s}\sum_{j\in \Gamma _{s-1}^{+}\cup \Gamma
_{s}^{-}}\widetilde{\delta }_{j}.
\]
Since r.v. $\{\widetilde{\delta }_{j}\}$ are independent of the trajectory $%
S_{k} $ and numbers $e^{s},$ $s=0,\pm 1,\pm 2,...$ , are rationally
independent, we have
\begin{eqnarray*}
P\{g_{2n} &=&I\}\sim \frac{1}{\sqrt{\pi n}}E\mathrel{\mathop{%
\stackrel{M_{2n}}{\Pi }}\limits_{s=m_{2n}+1}}\left(
\begin{array}{c}
2\tau _{s}^{-} \\
\tau _{s}^{-}
\end{array}
\right) (\frac{1}{2})^{2\tau _{s}^{-}}\leq \frac{1}{\sqrt{\pi n}}(\frac{1}{2}%
)^{M_{2n}-m_{2n}} \\
&=&\frac{1}{\sqrt{\pi n}}(\frac{1}{2})^{M_{2n}-m_{2n}}[I_{M_{2n}-m_{2n}>%
\sqrt{2n}}+I_{M_{2n}-m_{2n}<\sqrt{2n}}] \\
&\leq &\frac{1}{\sqrt{\pi n}}(\frac{1}{2})^{\sqrt{2n}}+\sum_{r=1}^{\sqrt{2n}%
}(\frac{1}{2})^{r}P\{|S_{k}|\leq r,\text{ }k=1,2,...2n,\text{ }S_{2n}=0\} \\
&\leq &e^{-c_{1}\sqrt{2n}}+\sum_{r=1}^{\sqrt{2n}}(\frac{1}{2}%
)^{r}P\{|S_{k}|\leq r,\text{ }k=1,2,...2n,\text{ }S_{2n}=0\}.
\end{eqnarray*}

\begin{lemma}
$P\{|S_{k}|\leq r,$ $k=1,2,...2n,$ $S_{2n}=0\}\leq (\cos \frac{\pi }{2(r+1)}%
)^{2n}.$
\end{lemma}

\textbf{Proof.} Let us introduce the operator $H_{0}\psi (x)=\frac{\psi
(x+1)+\psi (x-1)}{2}$ on the set $[-r,r]\in Z^{1}$\ with the Dirichlet
boundary conditions $\psi (r+1)=\psi (-r-1)=0.$ Then $\varphi (x)=\cos \frac{%
\pi x}{2(r+1)}$ is an eigenfunction of $H_{0}$ with the eigenvalue $\lambda
_{0,r+1}=\cos \frac{\pi }{2(r+1)}.$ Hence
\[
H_{0}^{2n}\varphi (x)=\lambda _{0,r+1}^{2n}\varphi (x).
\]
Let $p_{r}(k,x,z)$ be the transition probability of the random walk on $%
[-r,r]\in Z^{1}$\ with the absorption at $\pm (r+1).$\ Then
\[
\sum_{|z|\leq r}p_{r}(2n,x,z)\varphi (z)=\lambda _{0,r+1}^{2n}\varphi (x).
\]
Since $\varphi (z)\leq 1,$ $\varphi (0)=1,$ the latter relation implies
\[
\sum_{|z|\leq r}p_{r}(2n,x,z)\leq \lambda _{0,r+1}^{2n}.
\]
Since $S_{k},$ $k=0,1,...2n,$\ is the symmetric random walk on $Z^{1},$\ we
have
\[
P\{|S_{k}|\leq r,\text{ }k=1,2,...2n,\text{ }S_{2n}=0\}=p_{r}(2n,0,0)\leq
\lambda _{0,r+1}^{2n}.
\]

Direct calculation shows that
\[
\mathrel{\mathop{\max }\limits_{r\leq \sqrt{2n}}}(\frac{1}{2})^{r}(\cos
\frac{\pi }{2(r+1)})^{2n}\leq e^{-c(2n)^{1/3}},
\]
with the maximum achieved at $r=r_{0}\sim c_{1}(2n)^{1/3}.$ Thus
\[
P\{g_{2n}=I\}\leq (\frac{1}{2})^{\sqrt{2n}}+\sqrt{2n}e^{-c_{0}(2n)^{1/3}}%
\leq e^{-\widetilde{c}_{0}(2n)^{1/3}}
\]
for arbitrary $\widetilde{c}_{0}<c_{0}$ and sufficiently large $n.$ This
proves the first statement of the theorem. Now the second statement
follows from Theorem \ref{tll}.\qed

\bigskip

\textbf{Appendix. } \textbf{Proof of Theorem \ref{tael}.} As it was
mentioned after the statement of the theorem, it is enough to show the
validity of condition (b) and evaluate $\alpha ,\beta $. Let
\[
u_{t}=-H_{0}u,~t>0,~~~u|_{t=0}=f,
\]
with a compactly supported $f$ and
\[
\varphi =\varphi(x,\lambda) =\int_{0}^{\infty }ue^{\lambda t}dt,~~\mathrm{Re}\lambda \leq -a<0,~~x\in\Gamma^d.
\]
Note that we replaced $-\lambda $ by $\lambda $ in the Laplace transform
above. It is convenient for future notations. Then $\varphi $ satisfies the
equation
\begin{equation}
(H_{0}-\lambda )\varphi =f,  \label{res}
\end{equation}
and $u$ can be found using the inverse Laplace transform
\begin{equation}
u=\frac{1}{(2\pi )^{d}}\int_{-a-i\infty }^{-a+i\infty }\varphi e^{-\lambda
t}d\lambda .  \label{222}
\end{equation}
The spectrum of $H_{0}$ is $[0,\infty ),$ and $\varphi $ is analytic in $%
\lambda $ when $\lambda \in C\backslash \lbrack 0,\infty ).$ We are going to
study the properties of $\varphi $ when $\lambda \rightarrow 0$\ and $%
\lambda \rightarrow \infty .$ Let $\psi (z)=\psi(z,\lambda ),$ $z\in
Z^{d},$ be the restriction of the function $\varphi(x,\lambda),~x\in \Gamma^d, $ on the lattice $%
Z^{d}. $ Let $e$ be an arbitrary edge of $\Gamma ^{d}$ with end points $%
z_{1},z_{2}\in Z^{d}$ and parametrization from $z_{1}$ to $z_{2}$. By
solving the boundary value problem on $e$, we can represent $\varphi $ on $e$
in the form
\begin{equation}
\varphi =\frac{\psi (z_{1})\sin k(1-s)+\psi (z_{2})\sin ks}{\sin k}+\varphi
_{par},~~~\varphi _{par}=\int_{0}^{1}G(s,t)f(t)dt,  \label{repr}
\end{equation}
where $k=\sqrt{\lambda },$ Im$k>0,$ and
\[
G=\frac{1}{k\sin k}\left\{
\begin{array}{c}
\sin ks\sin k(1-t),~~s<t \\
\sin kt\sin k(1-s),~~s\geq t
\end{array}
\right. .
\]
Due to the invariance of $H_{0}$ with respect to translations and rotations
in $Z^{d}$, it is enough to estimate $p_{0}(t,x,x)$ when $x$ belongs to the
edge $e_{0}$ with $z_{1}$ being the origin in $Z^{d}$ and $%
z_{2}=(1,0,...,0). $ Let $f$ be supported on one edge $e_{0}.$ Then (\ref
{repr}) is still valid, but $\varphi _{par}=0$ on all the edges except $%
e_{0}.$ We substitute (\ref{repr}) into (\ref{sm}) and get the following
equation for $\psi :$%
\[
(\Delta -2d\cos k)\psi (z)=\frac{1}{k}\int_{0}^{1}\sin k(1-t)f(t)dt\delta
_{1}+\frac{1}{k}\int_{0}^{1}\sin ktf(t)dt\delta _{0},~~z\in Z^{d}.
\]
Here $\Delta $ is the lattice Laplacian defined in (\ref{lo}) and $\delta
_{0},$ $\delta _{1}$ are functions on $Z^{d}$ equal to one at $z,$ $y$,
respectively, and equal to zero elsewhere. In particular, if $f$ is the
delta function at a point $s$ of the edge $e_{0}$, then
\begin{equation}
(\Delta -2d\cos k)\psi =\frac{1}{k}\sin k(1-s)\delta _{1}+\frac{1}{k}\sin
ks\delta _{0}.  \label{1212}
\end{equation}

Let $R_{\mu }(z-z_{0})$ be the kernel of the resolvent $(\Delta -\mu )^{-1}$
of the lattice Laplacian. Then (\ref{1212}) implies that
\begin{equation}
\psi (z)=\frac{1}{\sqrt{\lambda }}\sin \sqrt{\lambda }sR_{\mu }(z)+\frac{1}{%
\sqrt{\lambda }}\sin \sqrt{\lambda }(1-s)R_{\mu }(z-z_{2}),~~\mu =2d\cos
\sqrt{\lambda }.  \label{asps}
\end{equation}

Function $R_{\mu }(z)$ has the form
\[
R_{\mu }(z)=\int_{T}\frac{e^{i(\sigma ,z)}d\sigma }{(\sum_{1\leq j\leq
d}2\cos \sigma _{j})-\mu },~~T=[-\pi ,\pi ]^{d}.
\]
Hence, function $~\sin (\sqrt{\lambda }s)R_{\mu }(z),~$ $s\in (0,1),~$ $\mu
=2d\cos \sqrt{\lambda },~$ decays exponentially as \\ $|$Im$\sqrt{\lambda }%
|\rightarrow \infty .$ This allows one to change the contour of integration in
(\ref{222}), when $z\in Z^{d},$ and rewrite (\ref{222}) in the form
\begin{equation}
u(z,t)=\frac{1}{(2\pi )^{d}}\int_{l}\psi _{\lambda }(z)e^{-\lambda
t}d\lambda ,~~z\in Z^{d},  \label{ll}
\end{equation}
where contour $l$ consists of the ray $\lambda =\rho e^{-i\pi /4},~\rho \in (\infty , 1)$, a smooth arc starting at $\lambda =e^{-\pi /4}$, ending at $\lambda =e^{\pi /4}$, and crossing the
real axis at $\lambda =-a$, and the ray $\lambda =\rho e^{i\pi /4},~\rho \in (1, \infty )$.
It is easy to see that $|\psi  (z,\lambda)|\leq C/|\sqrt{\lambda }|$ as $%
\lambda \in l$ uniformly in $s$ and $z\in Z^{d}.$ This immediately implies
that $|u(z,t)|\leq C/\sqrt{t}.$ Now from (\ref{repr}) it follows that the
same estimate is valid for $p_{0}(t,x,x),$ $x\in e_{0},$ i.e., condition (b)
holds, and $\alpha =1.$

From (\ref{ll}) it also follows that the asymptotic behavior of $u$ as $%
t\rightarrow \infty $ is determined by the asymptotic expansion of $\psi  (z,\lambda)$ as $\lambda \rightarrow 0,$ $\lambda \notin \lbrack 0,\infty
)$. Note that the spectrum of the difference Laplacian is $[-2d,2d],$ and $%
\mu =2d-d\lambda +O(\lambda ^{2})$ as $\lambda \rightarrow 0.$ From here and the
well known expansions of the resolvent of the difference Laplacian near the
edge of the spectrum it follows that the first singular term in the
asymptotic expansion of $R_{\mu }(z)$ as $\lambda \rightarrow 0,$ $\lambda
\notin \lbrack 0,\infty )$, has the form
\[
\left\{
\begin{array}{c}
c_{d}\lambda ^{d/2-1}(1+O(\lambda )),~~d~\mathrm{is}~\mathrm{odd,} \\
c_{d}\lambda ^{d/2-1}\ln \lambda (1+O(\lambda )),~~d~\mathrm{is}~\mathrm{even.%
}
\end{array}
\right.
\]
Then (\ref{asps}) implies that a similar expansion is valid for $\psi  (z,\lambda)$ with the main term independent of $s$ and the remainder
estimated uniformly in $s$. This allows one to replace $l$ in (\ref{ll}) by
the contour which consists of the rays $\arg \lambda =\pm \pi /4.$ From
here\ it follows that for each $z\in Z^{d}$ and uniformly in $s$,
\[
u(z,t)\sim t^{-d/2},~~t\rightarrow \infty .
\]
This and (\ref{repr}) imply the same behavior for $p_{0}(t,x,x),$ $x\in
e_{0},$ i.e., $\beta =d.$
\qed

\end{document}